\documentclass[12pt,indent]{article}
\usepackage{a4,latexsym,amsmath}
\usepackage[latin1]{inputenc}
\usepackage{float}
\usepackage{natbib}
\usepackage{graphics}
\usepackage[english]{babel}
\usepackage{tabularx}
\usepackage{lscape}
\usepackage{multirow}
\usepackage{rotating}
\usepackage{dsfont}
\usepackage{subfig}
\usepackage{bbm}
\usepackage{booktabs}
\usepackage[table]{xcolor}

\usepackage{calc}
\usepackage{ifthen}
\newenvironment{tabularsmall}
{ \footnotesize \sffamily \tabular } {
\endtabular
\normalfont }

\newcommand{\betab}{{\boldsymbol{\beta}}}

\newcommand{\deltab}{\boldsymbol{\delta}}

\newcommand{\gammab}{\boldsymbol{\gamma}}
\newcommand{\pib}{{\boldsymbol{\pi}}}

\newcommand{\thetab}{\boldsymbol{\theta}}

\newcommand{\xb}{\boldsymbol{x}}

\newcommand{\zb}{\boldsymbol{z}}

\newcommand{\0}{\boldsymbol{0}}
\newcommand{\1}{\boldsymbol{1}}

\newcommand{\blanco}[1]{}

\def\d{\displaystyle}

\usepackage{natbib}

\usepackage{setspace}

\title{DIFtreePCM}

\begin{document}
\bibliographystyle{chicago}
\sloppy

\makeatletter
\renewcommand{\section}{\@startsection{section}{1}{\z@}%
        {-3.5ex \@plus -1ex \@minus -.2ex}%
        {1.5ex \@plus.2ex}%
        {\reset@font\Large\sffamily}}
\renewcommand{\subsection}{\@startsection{subsection}{1}{\z@}%
        {-3.25ex \@plus -1ex \@minus -.2ex}%
        {1.1ex \@plus.2ex}%
        {\reset@font\large\sffamily\flushleft}}
\renewcommand{\subsubsection}{\@startsection{subsubsection}{1}{\z@}%
        {-3.25ex \@plus -1ex \@minus -.2ex}%
        {1.1ex \@plus.2ex}%
        {\reset@font\normalsize\sffamily\flushleft}}
\makeatother



\newsavebox{\tempbox}
\newlength{\linelength}
\setlength{\linelength}{\linewidth-10mm} \makeatletter
\renewcommand{\@makecaption}[2]
{
  \renewcommand{\baselinestretch}{1.1} \normalsize\small
  \vspace{5mm}
  \sbox{\tempbox}{#1: #2}
  \ifthenelse{\lengthtest{\wd\tempbox>\linelength}}
  {\noindent\hspace*{4mm}\parbox{\linewidth-10mm}{\sc#1: \sl#2\par}}
  {\begin{center}\sc#1: \sl#2\par\end{center}}
}



\def\R{\mathchoice{ \hbox{${\rm I}\!{\rm R}$} }
                   { \hbox{${\rm I}\!{\rm R}$} }
                   { \hbox{$ \scriptstyle  {\rm I}\!{\rm R}$} }
                   { \hbox{$ \scriptscriptstyle  {\rm I}\!{\rm R}$} }  }

\def\N{\mathchoice{ \hbox{${\rm I}\!{\rm N}$} }
                   { \hbox{${\rm I}\!{\rm N}$} }
                   { \hbox{$ \scriptstyle  {\rm I}\!{\rm N}$} }
                   { \hbox{$ \scriptscriptstyle  {\rm I}\!{\rm N}$} }  }

\def\d{\displaystyle}

\title{Item-Focussed Trees for the Detection of Differential Item Functioning in Partial Credit Models }
\author{Stella Bollmann, Moritz Berger \& Gerhard Tutz   \\{\small Ludwig-Maximilians-Universit\"{a}t M\"{u}nchen}\\
{\small Akademiestra{\ss}e 1, 80799 M\"{u}nchen}}


\maketitle

\begin{abstract} 
\noindent
Various methods to detect differential item functioning (DIF) in item response models are available. However, most of the methods assume that the responses are binary,  for ordered response categories available methods are scarce. In the present paper DIF in the widely used partial credit model is investigated. An item-focussed tree is proposed that allows to detect DIF-items, which might affect the performance of the partial credit model. The method uses tree methodology yielding a tree for each item that is detected as DIF-item. The resulting trees show  which variables induce DIF and in which way. The visualization as trees makes the results easily accessible. The method is compared to an alternative approach, simulations demonstrate the performance of the method and an application illustrates how it works for real data.
\end{abstract}


\section{Introduction}\label{sec:introduction}
For proper measurement, psychometric test models generally assume that test
and measurement properties are stable across individuals, stability  is also known
as measurement invariance \citep{millsap2012statistical}. However, it might occur
that different groups of people react differently on the same test and
validity of  measurements  is threatened. Also, test fairness
is violated if tests lead to different conclusions for different
groups of people.
When measurement invariance is violated on the item level it is called item
bias or differential item functioning (DIF). DIF is present if one ore more  items are
significantly more difficult for one group than for the other after controlling for
the underlying ability or trait. If the difference between the groups is constant
across different levels of ability or trait of the individual it is called \textit{uniform
DIF}. If this difference between groups is dependent on the ability or trait of
the person it is called \textit{non-uniform DIF}.
DIF detection procedures can also be classified into IRT methods and non-
IRT methods. The IRT methods, also called parametric methods, are those in
which an item response theory (IRT) model is used for the detection of DIF.
For an overview of IRT methods and non-IRT methods, see \citet{magisetal:2010} and \citet{wainer1993differential}.

The basic idea of traditional DIF detection procedures in both dichotomous and polytomous IRT models is to pre-specify two groups of persons and  then determine if item parameter estimates differ between these groups.
The first method that was introduced for the detection of DIF in IRT models is the Likelihood Ratio test (LRT; \citealp{Andersen:73}).
Another approach that can be used for any kind of IRT models is \emph{Lord's chi square test} \citep{Lord:1980}. While this test is restricted to the comparison of two groups, its extension by \citet{kim1995}, the \emph{generalized Lord test}, can be used for more than one focal group.  	
A third approach is the \emph{Raju} method \citep{raju1988area} that is based on the idea that the difference between the shape of item response curves (IRCs) between two groups indicates DIF.
Further test statistics to test for parameter differences between pre-specified groups were suggested by \citet{thissen1993detection}
and \citet{holland1988}. All of these classical methods have in common that they are limited to few sub-groups and these sub-groups have to be pre-specified by the user. Moreover, it is hard to consider more than one DIF inducing covariate at a time.

More recently two strategies were proposed that are able to detect DIF in Rasch models that is generated by multiple covariates and for which sub-groups do not have to be pre-specified.
The first strategy uses regularization methods to handle the abundance
of parameters in the model. \citet{TuSchauDIFPsych}, \citet{magis2014detection} and \citet{thissen1993detection} used penalized likelihood estimation whereas \citet{schauberger2015detection}
proposed boosting methods to obtain regularized estimates. The second strategy is to use  recursive partitioning techniques, often called tree methods. One has to distinguish between two quite different forms of tree methods in DIF detection. In the method proposed by
\citet{Stretal:2013:raschtree}, called RaschTree,  the covariate space is recursively partitioned
to identify regions of the covariate space in which item parameters differ. In the investigated regions a parametric latent trait model that includes covariates is fitted. Regions are suspected to be relevant if the parameter estimates in the regions differ strongly. Therefore, regions in the covariate space are identified that show different difficulties. A disadvantage of the method is that it detects regions of the covariate space that are linked to DIF but does not automatically detect the items that are responsible.
The alternative recursive partitioning method propagated by \citet{tutz2015item} focuses on the detection of the items that are responsible for DIF. Recursive partitioning is used on the item level not on the global level. In contrast to the RaschTree  it directly identifies  items that carry DIF. Since the method is able to flag DIF items it is referred to as item-focussed trees (IFTs).

For the partial credit model not many methods to detect DIF are yet available. An exception is \cite{el2014detecting}, in which the RaschTree approach has been extended to the multi-categorical case. The objective of the present paper is the development of item-focussed trees for the partial credit model.
In Section 2, the basic model and the used notation will be introduced. In addition we present an illustrative example. The tree algorithm that is used is given in detail in Section 3. In Section 4 we give   results of wider simulation studies. Finally, in Section 5 the new approach will be applied to an example of real data.

\section{DIF in Partial Credit Models}\label{sec:DIf}
In the following we consider $I$ items with ordered categories and $P$ persons. For simplicity we assume that the number of categories $k$ is equal across items.

\subsection{The Partial Credit  Model}
Let $Y_{pi} \in \{0,1,\dots,k\}$, $p=1,\dots,P$, $i=1,\dots,I$, denote the ordinal response of person $p$ on item $i$. The partial credit model (PCM), which was proposed by \citet{Masters:82}, assumes for the probabilities
\begin{equation}
\label{eq:PCM}
P(Y_{pi}=r)= \frac{\exp(\sum_{l=1}^{r}\theta_p-\delta_{il})}{\sum_{s=0}^{k_i}\exp(\sum_{l=1}^{s}\theta_p-\delta_{il})}, \quad r=1,\dots,k,
\end{equation}
where $\theta_p$ is the person parameter and $(\delta_{il},\dots,\delta_{ik})$ are the item parameters of item $i$.
For notational convenience   the definition of the model uses implicitly $\sum_{k=1}^{0}\theta_p-\delta_{ik}=0$.
With this convention an alternative form of the model is
\[
P(Y_{pi}=r)= \frac{\exp(r\theta_p-\sum_{k=1}^{r}\delta_{ik})}{\sum_{s=0}^{k_i}\exp(\sum_{k=1}^{s}\theta_p-\delta_{ik})}.
\]
The link to the binary Rasch model becomes obvious if one considers responses in adjacent categories. Given response categories $r$ and $r-1$, the presentation
\begin{equation}\label{eq:PCM11}
\log (\frac {P(Y_{pi}=r)}{P(Y_{pi}=r-1)})= \theta_p-\delta_{ir}, \quad r=1,\dots,k,
\end{equation}
shows that the model is locally a binary Rasch model with person parameter $\theta_p$ and item difficulty $\delta_{ir}$. The properties of the model can be visualized by item response curves (IRCs), which show the probabilities of a response in category $r$ as a function of the person parameter $\theta_p$.

\begin{figure}[!t]
		\centering
		\includegraphics[width=9cm]{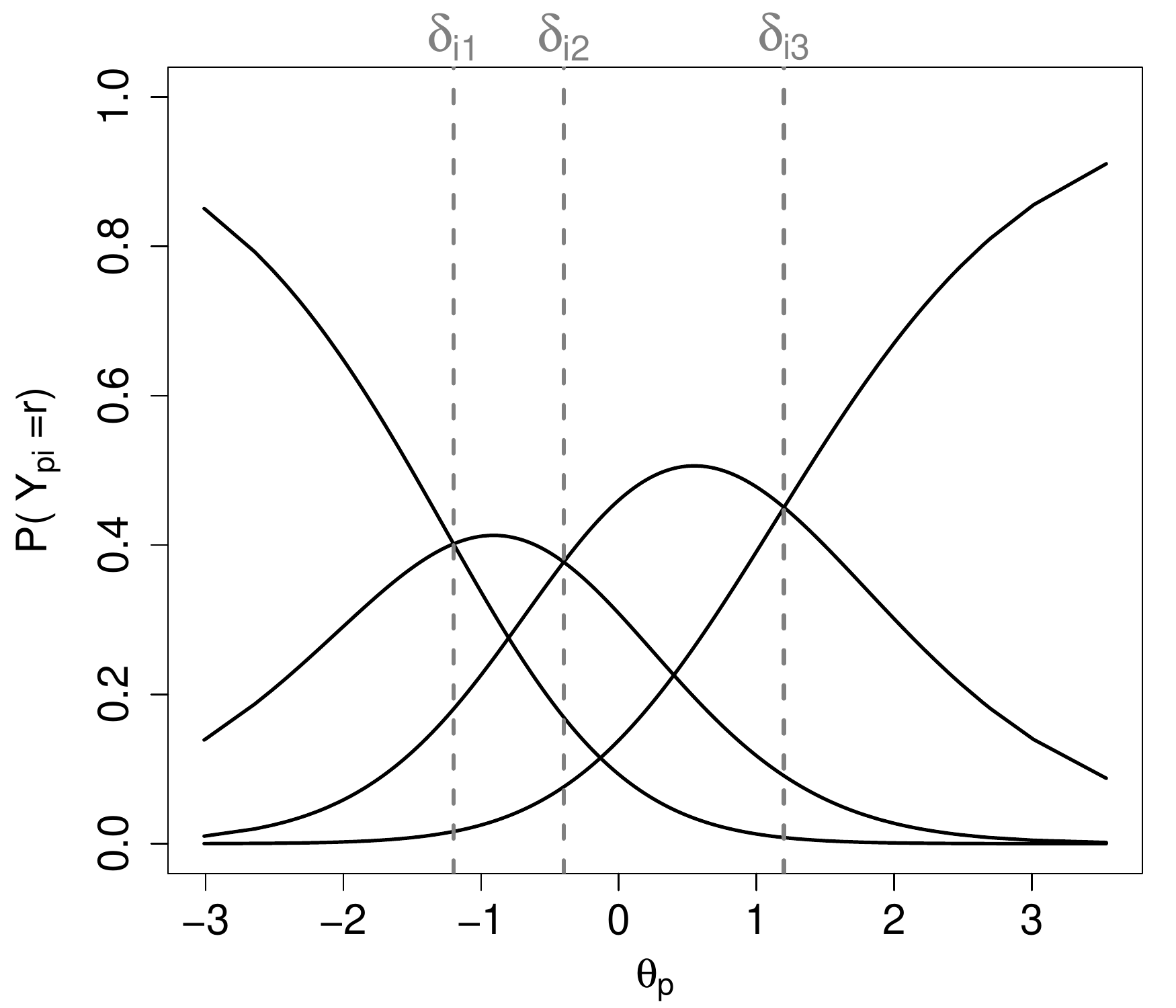}
		\caption{Item response functions (IRCs) for one item with four categories. The item parameters are marked by dashed lines.}
		\label{irf}
	\end{figure}

An example of the IRCs for one item with four categories is displayed in Figure \ref{irf}.
From the curves it is immediately seen that  for $\theta_p=\delta_{ir}$ the probabilities of adjacent categories are equal, that is, $P(Y_{pi}=r)={P(Y_{pi}=r-1)}$. That means the item response curves of adjacent categories intersect at $\theta_p=\delta_{ir}$. Therefore the parameters $\delta_{ir}$ can be seen as thresholds between categories $r-1$ and $r$. In Figure \ref{irf} the thresholds are marked by the dashed lines at the intersections of the curves. For example $Y_{pi}=0$ means that category 0 was chosen and no threshold was exceeded. The score $Y_{pi}=2$ implies a response which exceeds thresholds 1 and 2 but fails threshold 3. For more details of the model see also \citet{Masters:82}, \citet{MasWri:84} and \citet{Andrich:78,Andrich:13,Andrich:15}.

\subsection{Item-Focussed Trees for the Partial Credit  Model}
In representation (\ref{eq:PCM11}) the linear predictor for person $p$  and the $r$-th threshold of item $i$ is given by
\[
\eta_{pir}= \theta_p-\delta_{ir}.
\]
In item-focussed trees the predictor is successively modified by allowing different predictors in different regions of the covariate space. In the simple  case of a continuous variable $x$ one allows that the region is split into the region $\{x \le c\}$ and $\{x > c\}$ at split-point $c$. A tree is grown by successive splitting of one of the available variables at one of the corresponding split-points. The root is the top node without splitting, the terminal nodes represent the identified partitioning of the covariate space.

For a more concise description, let $\xb_p^T=(x_{p1},\dots,x_{pV})$ denote a vector of measurements on person $p$.
Starting from the root, the predictor that is fitted for item $i$ and all persons has the form
\[
\eta_{pir} = \theta_p - [\gamma_{ir(1)} I(x_{pv} \le c_{v})+ \gamma_{ir(2)} I(x_{pv} > c_{v})],\quad r=1,\dots,k,
\]
where $I(\cdot)$ denotes the indicator function with $I(a)=1$ if $a$ is true and $I(a)=0$ otherwise. That means, item $i$ shows DIF generated by the $v$-th variable. The item has parameters
$\gamma_{i1(1)},\dots, \gamma_{ik(1)}$ in the left node $I(x_{pv} \le c_{v})$ and parameters $\gamma_{i1(2)},\dots, \gamma_{ik(2)}$ in
the right node $I(x_{pv} > c_{v})$. The split-point $c_{v}$ defines the regions that are used and has to be chosen appropriately.

Further splitting means that one of the nodes, for example the left node $I(x_{pv} \le c_{v})$,  is further split in variable $s$, yielding the partition into left and right node
\[
I(x_{pv} \le c_{v})I(x_{ps} \le c_{s}) \quad \text{and} \quad I(x_{pv} \le c_{v})I(x_{ps} > c_{s}),
\]
where $c_{s}$ is a new split point for variable $x_{ps}$. For each region one again obtains new parameters for the item.
Of course, only items should be split that carry DIF and the variables and their split-points have to be selected carefully.

In the following we use the model abbreviation PCM-IFT for item-focussed trees based on the PCM.

\subsection{An Illustrative Example}
\label{sec:ill.ex}

\begin{figure}[!t]
			\centering
			\includegraphics[width=\textwidth,height=7cm]{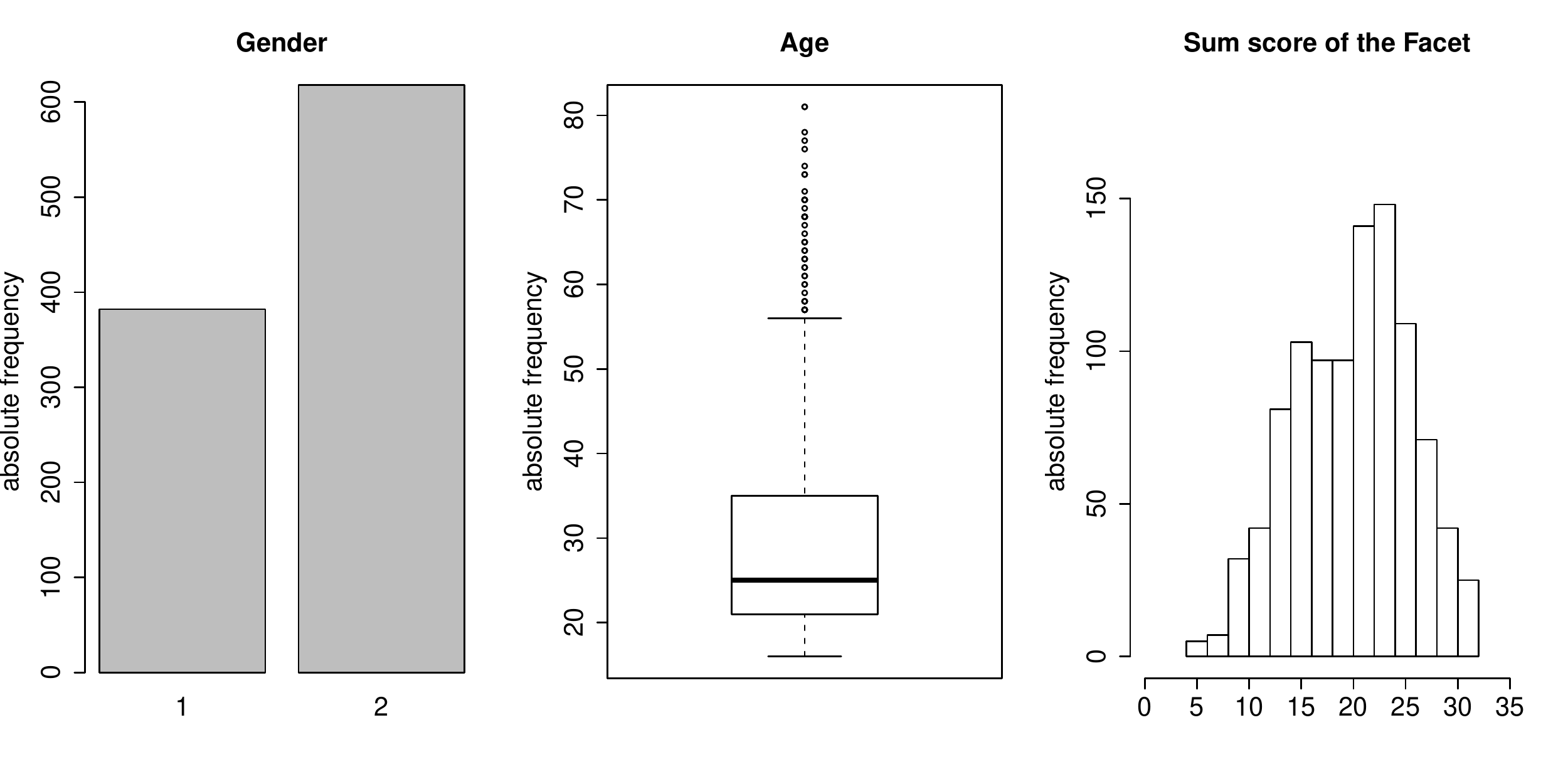}
			\caption{Graphical representation of the distribution of the sum score of the facet \emph{Achievement striving} and the two covariates (NEO-PI-R).}
			\label{descriptives_C4}
		\end{figure}

Before giving the fitting procedure of the proposed model in detail (see Section \ref{sec:Fit}) we consider an illustrative example.
The data considered here are the responses of 1000 subjects on the 8 items of the sub-facet \emph{Achievement striving} of the factor \emph{Conscientiousness} of the German version of the NEO personality inventory revised (NEO-PI-R; \citealp{ostendorf_neo-personlichkeitsinventar_2004}). The 1000 subjects were randomly drawn out of the 11,724 cases of the norm data set. The sample was taken for obtaining standard values for the test manual. Each of the items has five categories. Additionally, the data set comprises the two variables age and gender. The distribution of the sum score of the sub-facet and the covariates are shown in Figure \ref{descriptives_C4}.

The major domain \emph{Conscientiousness} is described in the manual as \emph{degree of organization, persistence, control and motivation in goal directed behaviour} and the sub-facet \emph{Achievement striving} as \emph{need for personal achievement and sense of direction} \citep{ostendorf_neo-personlichkeitsinventar_2004}.

\begin{figure}[!t]
 			\centering
 			{\includegraphics[width=\textwidth]{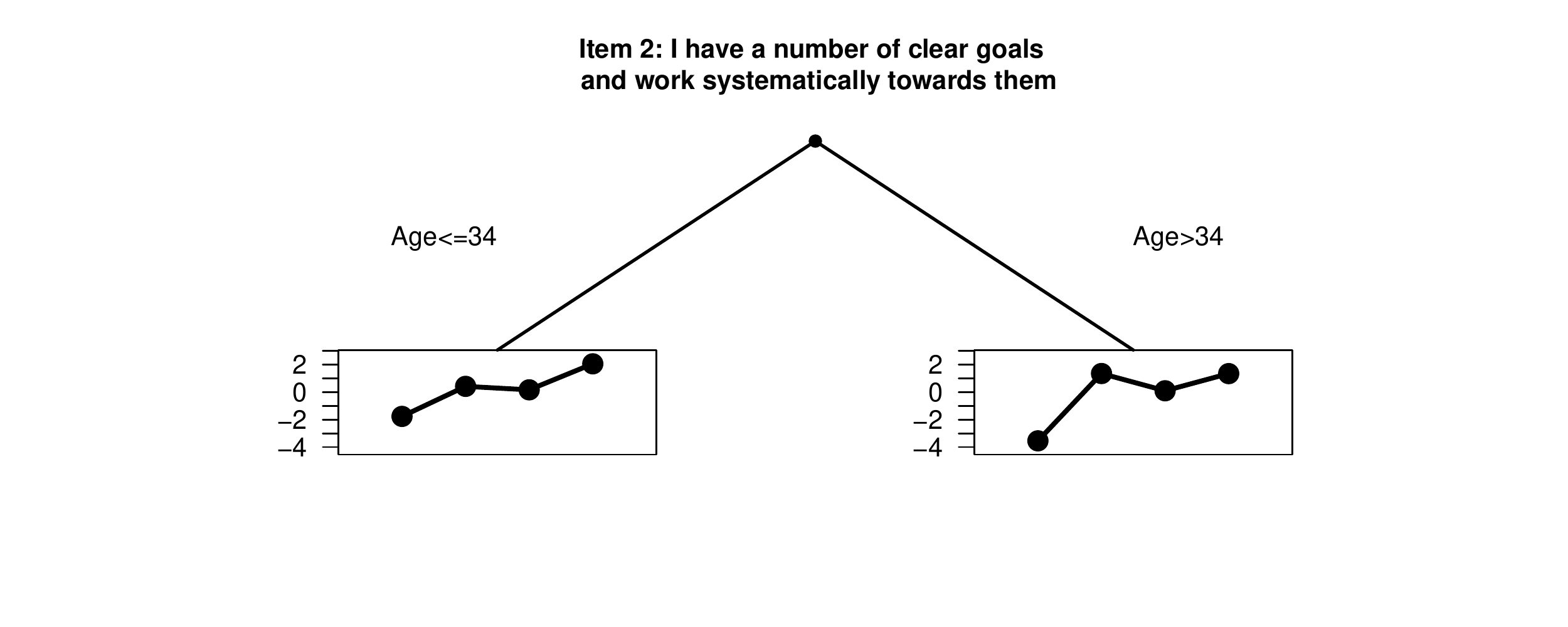}}
 	
 	
 			{\includegraphics[width=\textwidth]{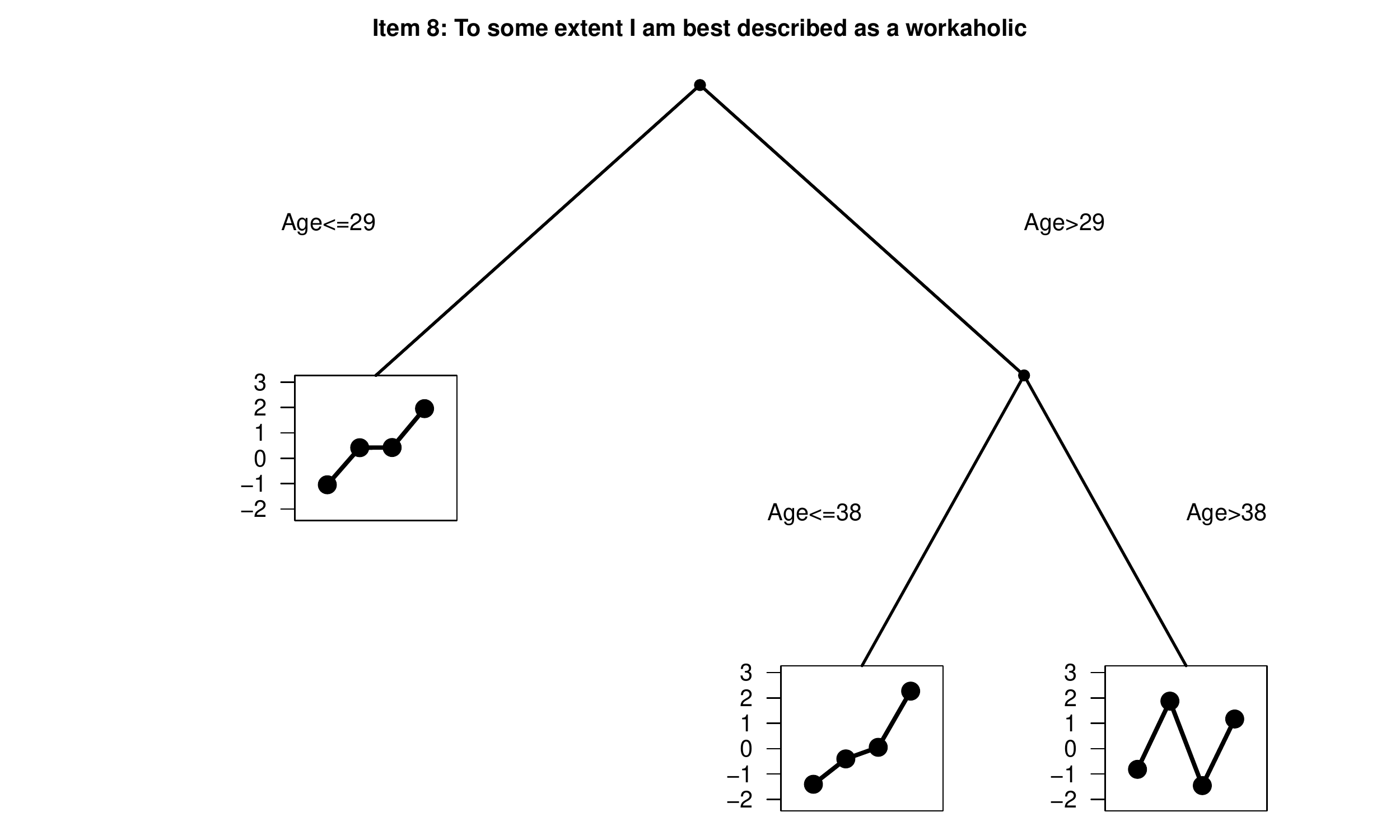}}
 	\caption{Trees for item 2 and item 8 of the sub-facet \emph{Achievement striving} (NEO-PI-R).}
 			\label{trees_C4}
 		\end{figure}

Using PCM-IFT, two of the eight items were detected as DIF items. The two items are the following:

\begin{itemize}
\item[] Item 2: I have a number of goals and work systematically towards them.
\item[] Item 8: To some extent I am best described as a workaholic.
\end{itemize}
Both items were only split in covariate age, but no significant split was found for covariate gender. The algorithm performs three splits until further splits are not significant anymore at a significance level of $\alpha=0.05$ (for further details of the test see Section \ref{sec:Fit}). Item 2 was split once and item 8 was split twice. The resulting trees for the two items are shown in Figure \ref{trees_C4}. At each terminal node of the trees the four estimated item parameters are shown in a graphical representation. It can be seen that all the item parameters are allowed to vary freely within the groups defined by the executed splits.

For item 2, the main difference between the two groups is that the first threshold $\delta_{21}$ is lower and the second threshold $\delta_{22}$ is higher for persons older than $34$. This means that more people chose the second category compared to the first category. Furthermore, in both groups the thresholds are not ordered. This effect is more extreme for older persons ($Age>34$). One reason might be that very few people chose the middle category in this group.

For item 8 one has to distinguish between young people ($Age\leq29$), middle-aged people ($30<Age\leq 38$) and older people ($Age>38$). In the latter group a severe violation of the ordering of categories can be observed. It seems that in this group a comparatively low latent trait was required to jump from the third to the fourth category. In fact, almost half of the people in this group chose the fourth category (45.18 \%).
 	
This illustration shows that the proposed PCM-IFT may be a useful tool for the detection of DIF in ordered items. The   performance of the new approach compared to an existing tree-based approach will be investigated in more detail in Section \ref{sec:sim} and Section \ref{sec:App}.
 		
%

\section{Fitting Item-Focussed Trees}\label{sec:Fit}
In this section we give a detailed description of the fitting procedure for the proposed PCM-IFT.

\subsection{The Partial Credit  Model as a Generalized Linear Model}

Under usual assumptions the partial credit model can be embedded into the framework of multivariate generalized linear models (GLM). Let the data be given by $(Y_{pi},\xb_p),\,p=1,\hdots,P,\,i=1,\hdots,I$. For the item responses one assumes a multinomial distribution $Y_{pi}|\xb_p \sim M(1,\pib_{pi})$, where $\pib_{pi}^\top=(\pi_{pi1},\hdots,\pi_{pik})$ with components $\pi_{pir}=P(Y_{pi}=r|\xb_p)$. The link function of the GLM can be derived from representation \eqref{eq:PCM11} and has the form
\begin{equation}\label{eq:PCM_GLM}
g(\pi_{pir})=\eta_{pir}=\log \left(\frac {P(Y_{pi}=r)}{P(Y_{pi}=r-1)}\right)= (\1_p^{(P)})^\top\thetab-(\1_r^{(k)})^\top\deltab_i,
\end{equation}
where $\thetab^\top=(\theta_1,\hdots,\theta_{P})$, $\deltab_{i}^\top=(\delta_{i1},\hdots,\delta_{ik})$ and $\1_r^{(k)}$ denotes the unit vector of length $k$ with a $1$ in component $r$. To ensure the identifiability of model \eqref{eq:PCM_GLM} one parameter has to be fixed. In the following we set $\theta_P=0$.
By defining the whole parameter vector $\betab^\top=(\thetab^\top,\deltab_1^\top,\hdots,\deltab_I^\top)$ the PCM can be written in the closed form
\[
\eta_{pir}=\zb_{pir}\betab,
\]
where $\zb_{pir}$ is the design vector for person $p$, item $i$ and threshold $r$ that has to be specified accordingly.

\subsection{Computation of Estimates} \label{computation}

Estimates of model \eqref{eq:PCM_GLM} can be obtained by use of the flexible R-package \texttt{VGAM} (\citealp{yee2010vgam}; \citealp{VGAM2014}). Function \texttt{vglm()} allows to estimate so-called vector generalized linear models \citep{YeeWil:96}. One just has to specify the design matrix as described above and estimation can easily be obtained. In addition one can make use of the argument \texttt{parallel()} to specify category-specific item parameters. In the following algorithm, which yields item-focussed trees based on the PCM, this estimation procedure serves as a building block in each iteration.

\subsection{Fitting of Trees}

When growing trees one has to take two decisions in each step. One has to determine the best split due to an optimality criterion and has to decide if the split is relevant or not. In contrast to alternative approaches the trees are not pruned to an adequate size after building an oversized tree. By early stopping   the size of the trees is controlled directly.

To determine the first split one examines for all the items, all the variables and possible split-points the PCM with predictors
\[
\eta_{pir} = \theta_p - [\gamma_{ir(1)} I(x_{pv} \le c_{v})+ \gamma_{ir(2)} I(x_{pv} > c_{v})],\quad r=1,\dots,k.
\]
DIF occurs, if $\gammab_{i(1)}\neq\gammab_{i(2)}$, where $\gammab_{i(\ell)}^T=(\gamma_{i1(\ell)},\dots,\gamma_{ik(\ell)})$, $\ell \in \{1,2\}$. The corresponding hypothesis $H_0:\gammab_{i(1)}-\gammab_{i(2)}=\0$ can be tested by a likelihood ratio (LR) test. One simply selects the combination of item, variable and split-point that yields the smallest $p$-value, which is equivalent to selecting the model with minimal deviance.
In later steps the basic procedure is the same. One performs LR-tests for the two parameter sets that are involved in the splitting and selects the combination  that yields the smallest $p$-value as the optimal one.

In order to determine the optimal size of the trees one has to decide in each step if the split should be performed or not. In answering this question one investigates the dependence of the response and the selected variable. For fixed item $i$ and variable $v$ let the maximal value statistic
$T_v=max_{c_v}T_{vc_v}$ be defined as the maximum of all the LR test statistics $T_{vc_v}$, where $c_v$ is from the set of possible split-points. Typically the test statistics $T_{vc_v}$ are strongly correlated. The relevance of variable $v$ is judged by the $p$-value of the distribution of $T_v$, which is not influenced by the number of split-points, since it is already taken into account, see \citet{HotLau:03}, \citet{Shih:04}, \citet{ShiTsa:2004}, \citet{StrBouAug:2007}. For the decision on the null hypothesis controlling for a given significance level $\alpha$ a permutation test is used. Thus, no distributional assumption has to be made. The test statistic $T_v$ is computed based on a data matrix in which variable $v$ is randomly permuted. The maximal value statistics for a large number of permutations provide a distribution of $T_v$ under the assumption of the null hypothesis that variable $v$ has no effect. The derived $p$-value is used to make the splitting decision.

Finally one has to address the problem of multiple testing. In DIF detection one typically controls for the type I error, that is, the item-wise significance level. To ensure that the proposed procedure also controls this level a Bonferroni adjustment is applied. For fixed item and variable the local significance level for one permutation test is set to $\alpha/V$, where V is the number of variables. Using this adaption  the probability of a false DIF result or the probability of falsely identifying at least one variable as responsible for DIF is controlled by $\alpha$. Of course the adjustment is only applied when several variables are available. If in later steps a variable is no longer available because all possible splits were already performed, the adaption consequently is changed to $V-1$ in all further nodes. All the results presented in this article are based on significance level $\alpha=0.05$ and 1000 permutations. This ensures that the $p$-values can be determined with sufficient accuracy.

A second criterion that is used to define the size of the trees is the minimal sample size in each node. In order to provide a sufficient basis for parameter estimation in each node, splitting is stopped in an item if the number of observations in any of the nodes falls below a predefined threshold. In our applications and simulations we used 30 observations. This value is in accordance to the choice of El-Komboz et. al. (2014).

If no further significant effect is found or splitting is stopped due to minimal node sizes the algorithm stops. After several splits each node can be represented by a product of B indicator functions, namely
\[
node(\xb_p)=\prod_{b=1}^{B} I(x_{pj_b} > c_{j_b})^{a_{b}}I(x_{pj_b} \leq c_{j_b})^{1-a_{b}},
\]
where $B$ is the total number of indicator functions or branches, $c_{j_b}$ is the selected split point in variable $j_b$ and $a_b \in \{0,1\}$ indicates which of the indicator functions, below or above the threshold, is involved.
Using this definition the final model of an item $i$ that has been split can be represented by
\[
\eta_{pir}=\theta_p + tr_{ir}(\xb_p)=\theta_p-\sum_{\ell=1}^{L_i}\gamma_{ir(\ell)}\,node_{i\ell}(\xb_p),\quad r=1,\dots,k,
\]
where $tr_{ir}(\xb_p)$ is the tree component containing sub-group specific threshold parameters $\gammab_{ir}$ and  $\ell=1,\hdots,L_i$ denote the terminal nodes of the tree. If an item is never chosen for splitting it is assumed to be free of DIF and the constant $tr_{ir}(\xb_p)=\delta_{ir}$, corresponding to the threshold of the simple PCM, is fitted.
\\[1em]
A concise description of the basic algorithm is given in the following.

\vspace{0.5cm}
\hrule
\begin{center}{\bf Basic Algorithm - PCM-IFT}\end{center}

\begin{description}
\item{\it Step 1 (Initialization)}

Set counter $m=1$

\begin{itemize}
\item[(a)] Estimation

For all items $i=1,\hdots,I$, fit all the candidate PCMs with predictors
\begin{align*}
\eta_{pir} = &\; \theta_p - [\gamma_{ir(1)} I(x_{pv} \le c_{vj})+ \gamma_{ir(2)} I(x_{pv} > c_{vj})],\\
& v=1,\hdots,V,\quad j=1,\hdots,J_v
\end{align*}

\item[(b)] Selection

Select the model that has the best fit. Let $c_{v_1,j_1}$ denote the best split, which is found for item $i_1$ and variable $x_{v_1}$.

\item[(c)] Splitting decision

Select the item and variable with the largest value of $T_v$. Carry out permutation test for this combination with significance level $\alpha/V$. If significant, fit the selected model yielding estimates $\hat{\thetab}_p$, $\hat{\gammab}_{i_1,1}$, $\hat{\gammab}_{i_1,2}$ and nodes $node_{i_1,1}, node_{i_1,2}$, set $m=2$. If not, stop, no DIF detected.
\end{itemize}

\item{\it Step 2 (Iteration)}

\begin{itemize}
\item[(a)] Estimation:

For all items $i=1,\hdots,I$ and already built nodes $\ell=1,\hdots,L_{im}$, fit all the candidate logistic models with new intercepts
\[
\gamma_{i,L_{im}+1}node_{i\ell}I(x_{pv}\leq c_{vj})+\gamma_{i,L_{im}+2}node_{i\ell}I(x_{pv}>c_{vj})
\]
for all v and remaining, possible split points $c_{vj}$.

\item[(b)] Selection

Select the model that has the best fit yielding the split point $c_{v_m,j_m}$, which is found for item $i_m$ in node $node_{i_m,\ell_m}$ and variable $x_{v_m}$

\item[(c)] Splitting decision

Select the node and variable with the largest value of $T_v$. Carry out permutation test for this combination with significance level $\alpha/V$. If significant, fit the selected model yielding the additional estimates $\hat{\gammab}_{i_m,L_{i_m,m}+1}, \hat{\gammab}_{i_m,L_{i_m,m}+2}$, set $m=m+1$. If not, stop.

\end{itemize}
\end{description}
\hrule
\vspace{0.5 cm}

\section{Simulation Studies}\label{sec:sim}

	In this section, we examine the performance of the new PCM-IFT approach that was introduced in the previous sections. More precisely we evaluate the procedure's ability to detect items that show DIF and to estimate the item difficulty parameters in each node in three   simulation studies. In addition, we compare the performance to the competitive approach proposed by \cite{el2014detecting}.
It is an  approach that fits IRT models separately in sub populations.  It is global in the sense that it looks for significant differences in parameter estimates in two different samples for \emph{all of the items}. The idea of the method is  to search for the split point with the highest parameter difference out of all possible split-points. However, the whole partial credit model is fitted separately in the sub populations.
	
In Simulation I (Section \ref{SimI}) a simple model with only one binary covariate will be considered. In Simulation II a more complex model with three different covariates (binary, ordinal and numeric) will be the data generating model (Section \ref{SimII}). Finally, in Simulation III (Section \ref{SimIII}) non-homogeneous DIF will be considered in a simulation with one binary covariate.

    	\subsection{Evaluation Criteria and Experimental Design}
	 For the evaluation of simulation results in each simulation scenario true positive rates (TPR) and false positive rates (FPR) are reported. 
	
    Let \emph{each item} be characterized by a vector $\boldsymbol{\epsilon}_i^T=(\epsilon_{i1},\dots,\epsilon_{iV})$ with $\epsilon_{iv}=1$ if item $i$ has DIF in variable $v$ and $\epsilon_{iv}=0$ otherwise. An item is a non DIF item if $\boldsymbol{\epsilon}_i^T=(0,\dots,0)$. As soon as one of the components is 1, it is a DIF item. In addition \emph{each variable} can be characterized by a vector $\boldsymbol{\epsilon}_v^T=(\epsilon_{v1},\dots,\epsilon_{vI})$, where $\epsilon_{vi}=1$ if variable $v$ induces DIF in item $i$ and $\epsilon_{vi}=0$ otherwise. With $\hat{\boldsymbol{\epsilon}}_i^T=(\hat{\epsilon}_{i1},\dots,\hat{\epsilon}_{iV})$ denoting the corresponding estimated indicator vector, the indicator function $I(\cdot)$ and the zero vector $\boldsymbol{0}$, the following criteria are used:
    	\begin{enumerate}
		\item TPR and FPR on the item level:
       \[
			TPR_I=\frac{1}{\#\{i:\boldsymbol{\epsilon}_i\neq \boldsymbol{0}\}}\sum_{i:\boldsymbol{\epsilon}_i\neq0}I(\hat{\boldsymbol{\epsilon}}_i\neq \boldsymbol{0})
			\]
       \[
			FPR_I=\frac{1}{\#\{i:\boldsymbol{\epsilon}_i= \boldsymbol{0}\}}\sum_{i:\boldsymbol{\epsilon}_i=0}I(\hat{\boldsymbol{\epsilon}}_i\neq \boldsymbol{0})
			\]
			
       \item TPR and FPR for the combination of item and variable:
       \[
			TPR_{IV}=\frac{1}{\#\{i,v:\epsilon_{iv}\neq 0\}}\sum_{i,v:\epsilon_{i,v}\neq0}I(\hat{\epsilon}_{iv}\neq0)
			\]
       \[
			FPR_{IV}=\frac{1}{\#\{i:\epsilon_{iv}= 0\}}\sum_{i,v:\epsilon_{iv}=0}I(\hat{\epsilon}_{iv}\neq0)
			\]
       \item TPR and FPR on the variable level:

       \[TPR_V=\frac{1}{\#\{v:\boldsymbol{\epsilon}_v\neq \boldsymbol{0}\}}\sum_{v:\boldsymbol{\epsilon}_v\neq0}I(\hat{\boldsymbol{\epsilon}}_v\neq \boldsymbol{0})\]

       \[FPR_V=\frac{1}{\#\{v:\boldsymbol{\epsilon}_v = \boldsymbol{0}\}}\sum_{v:\boldsymbol{\epsilon}_v=0}I(\hat{\boldsymbol{\epsilon}}_v\neq \boldsymbol{0})\]
\end{enumerate}
Each rate is reported as the average over all repetitions. All simulation scenarios were replicated 50 times.


\paragraph{Person Parameters}
The number of persons in all simulations is 500. First, all persons are excluded, who have  answers in only  one category. As a result, the actual number of persons $P$ in most of the scenarios is slightly less then 500. The person parameters are simulated from a standard normal distribution, $\theta_p \sim N(0,1)$.



	
	\paragraph{Number of Items}
	In most scenarios the number of items is $I=8$, and one of these items is simulated to have DIF. This makes our simulations comparable to the real data examples in Section \ref{sec:ill.ex} and Section \ref{sec:App}, where each unidimensional sub-facet consists of 8 items. Also \cite{el2014detecting}   used 8 items in their simulation studies.
	In order to examine how the performance of our method changes with increasing number of items, in Simulation I we conduct one scenario with $I=20$ and three DIF items.
	
	\paragraph{Item Parameters}
	In most scenarios we simulate data with three response categories (k=2). In addition, in Simulation I one scenario is included with five response categories (k=4). In a first step, the threshold parameters for item $i$ are drawn from the following normal distribution:
\begin{align*}
&k=2: \boldsymbol{\delta}_i \sim N_3(\boldsymbol{\mu}_3,\boldsymbol{\Sigma}_3=\boldsymbol{I_3}), \quad  \boldsymbol{\mu_3} = (-0.50 , 0.50)^\top \\
&k=4: \boldsymbol{\delta}_i \sim N_5(\boldsymbol{\mu_5},\boldsymbol{\Sigma_5}=\boldsymbol{I_5}), \quad \boldsymbol{\mu_5} = (-1.50 , -0.50 , 0.50 , 1.50)^\top\\
\end{align*}
If item $i$ is simulated to have DIF the corresponding item parameters are subsequently transformed by step functions.

\paragraph{Structure of DIF}
To simulate DIF in item $i$, the item parameters are shifted for one sub-group (the focal group) corresponding to a pre-specified split-point $c_{vj}$ in covariate $x_{v}$. There is always one split in each DIF item.

For each scenario, we define three different strengths of DIF: weak, medium and strong. The strength is determined by an additional parameter $\lambda$. In the \emph{weak} condition the mean vector of the focal group is shifted by $\lambda=0.25$, in the \emph{medium} condition by $\lambda=0.5$ and in the \emph{strong} condition by $\lambda=1$ in relation to the values in the reference group. Additionally, we add one condition in which \emph{no DIF} is present (the item parameters for both groups are drawn from the same distribution). Further details are given in the respective sections.
\\[1em]
The methods considered in the simulations are:
    \begin{itemize}
    \item The proposed \emph{item-focussed tree approach} (PCM-IFT) that was described in the previous sections.
    \item The \emph{partial credit tree approach} (TREE-PCM) proposed by \cite{el2014detecting}.
    \end{itemize}
During estimation each permutation test is based on 1000 permutations and global significance level $\alpha=0.05$.

   \subsection{Simulation I: One Binary Covariate}
    \label{SimI}

	In the first simulation study the data set contains only one binary covariate $x \in \{0,1\}$. Covariate $x$ induces DIF in one or three items. The item parameters for the two groups defined by $x$ are
\[
\gamma_{ir(2)}=\gamma_{ir(1)}+\lambda \cdot I(x_{p}=1), \quad r=1, \dots, k.
\]
All thresholds of the DIF items are shifted in the same direction by the same value $\lambda$ depending on the strength of DIF. For the settings with no DIF $\lambda$ is set to $0$.

We consider three scenarios that differ with regard to the number of items ($I$) the number of response categories ($k$) and the number of DIF items ($I_{DIF}$). A detailed overview is given in Table \ref{Scenarios1}.
	
	\begin{table}[!t]
	\caption{Number of Items (I), number of response categories (k) and number of DIF items ($I_{DIF}$) for the three scenarios of Simulation I.}
	\begin{center}
	\begin{tabularsmall}{lrrr}
	\toprule
	Simulation I&I&k&$I_{DIF}$\\
	\midrule
	Scenario 1&8&3&1\\
	Scenario 2&20&3&3\\
	Scenario 3&8&5&1\\
	\bottomrule
	\end{tabularsmall}
	\label{Scenarios1}
	\end{center}
	\end{table}

	
%
%
%

        \subsubsection*{Results  }

    The evaluated criteria of the first simulation are shown in Table \ref{tab:1.simulation.rate.ift} and Table  \ref{tab:1.simulation.rate.alt}. For the case where no DIF is present, only false positive rates are available. In both tables first results for eight items with three categories are shown, second for 20 items with three categories and third for eight items with five categories. In the case of one single covariate the covariate vector $\epsilon_i$ only has one element, so true and false positive rates for the combination of item and variable for PCM-IFT correspond to those on the item level (see Table \ref{tab:1.simulation.rate.ift}). TREE-PCM does not test single items for DIF and therefore we only get the detection rates on the covariate level. In the no DIF scenario we get a $FPR_V$ and in all other scenarios a $TPR_V$. They are reported for both methods in Table \ref{tab:1.simulation.rate.alt}.

    \begin{table}[!t]
		\caption{True positive and false positive rates for PCM-IFT (Simulation I)}
		\begin{center}
			\begin{tabularsmall}{llcc}
				
				\toprule
       & \makebox{\textbf{DIF strength}}      & $\boldsymbol{TPR_I}$ & $\boldsymbol{FPR_I}$ \\
                \cline{2-4}
       Scenario 1 &     no DIF  & - & 0.058 \\
			&	weak  & 0.260 & 0.057 \\
			&	medium  & 0.820 & 0.057 \\
			&	strong  & 1.000 & 0.054 \\
            \cline{2-4}
	Scenario 2 &	no DIF  & - 	& 0.059 \\
			&		weak 	 & 0.240 & 0.058 \\
			&		medium  & 0.760 & 0.059 \\
			&		strong  & 0.980	& 0.055 \\
          \cline{2-4}
	Scenario 3 &no DIF  & - 	& 0.055 \\	
            &    weak  & 0.360 & 0.060 \\
			&	medium  & 0.920 & 0.063 \\
			&	strong  & 0.980 & 0.060 \\

				\bottomrule
			\end{tabularsmall}
		\end{center}
		
		\label{tab:1.simulation.rate.ift}
	\end{table}

It can be seen from Table \ref{tab:1.simulation.rate.ift} that PCM-IFT approximately keeps the given significance level. As was to be expected, true positive rates on the item level increase with increasing strength of DIF and they are also slightly higher for the third scenario in which items with 5 categories instead of 3 were simulated. The false positive rates on the variable level (Table \ref{tab:1.simulation.rate.alt}) seem surprisingly high. Bearing in mind though that false positive rates were controlled on the item and not on the variable level, the results  make  sense. If the probability of one item to be falsely classified as DIF item is 0.05, then the probability that one or more out of 8 items is falsely classified as DIF item is: $1-(0.95^8)=1-0.663=0.337$ and for 20 items: $1-(0.95^{20})=1-0.358=0.642$. Of course, this only holds for simulation I in which there is only one covariate and each split is automatically made for this covariate.
Consequently, false positive rates on the variable level are much higher compared to the TREE-PCM procedure, in which they are controlled on the variable level, and therefore the significance level is mostly respected. It can further be seen, that also true positive rates are much higher for PCM-IFT than for TREE-PCM. A true positive rate of 0.140 in scenario 3 with weak DIF means that only in 14 \% of the cases the present DIF is found. The reason might be that the ratio  of DIF items to non-DIF items is very small in scenarios 1 and 3. Therefore, for the detection of single items the power is much higher. Accordingly, in scenario 2 where the ratio of DIF items to non-DIF items is higher TREE-PCM performs better.

\begin{table}[!t]
		\caption{ $TPR_V$ and $FPR_V$ for TREE-PCM and PCM-IFT (Simulation I)}
		\begin{center}
			\begin{tabularsmall}{llcccc}
				
				\toprule
				& \makebox{\textbf{DIF strength}} & \multicolumn{2}{c}{\textbf{TREE-PCM}} & \multicolumn{2}{c}{\textbf{PCM-IFT}} \\
				&&$TPR_V$&$FPR_V$&$TPR_V$&$FPR_V$\\
                \cline{2-6}
     Scenario 1 &     no DIF &---	& 0.100 &---& 0.380  \\
			&	weak 		& 0.100 &---	& 0.480 &---   \\
			&	medium		 & 0.320 &---	& 0.860 &---   \\
			&	strong 		& 0.900 &---	& 1.000 &---   \\
            \cline{2-6}
	Scenario 2 &		    no DIF &---& 0.040 &---& 0.720  \\	
            &    weak & 0.220 &---& 0.820 &--- \\
			&	medium & 0.860 &---& 1.000 &--- \\
			&	strong & 1.000 &---& 1.000 &--- \\
            \cline{2-6}
	Scnenario 3 &no DIF &---	& 0.060 &---	& 0.444 \\
			&	weak 	& 0.140 &---	& 0.600 &--- \\
			&	medium & 0.580 &---		& 0.940 &--- \\
			&	strong & 1.000 &---		& 0.980 &--- \\
				\bottomrule

			\end{tabularsmall}
		\end{center}
		
		\label{tab:1.simulation.rate.alt}
	\end{table}

 \subsection{Simulation II: Three Different Covariates}
  \label{SimII}

  	In the second simulation study, we investigate how well the proposed method is able to detect the right DIF inducing covariate out of multiple present covariates. We consider scenarios with $I=8$, $k=3$ and $I_{DIF}=1$. Now, there are three different covariates that possibly induce DIF - one binary variable $x_1 \in \{0,1\}$, one ordered factor $x_2 \in \{1,2,3,4\}$ and one numeric covariate $x_3 \in \{20,\dots,50\}$. Variable $x_3$ could, for example, represent the variable age. In each of the following scenarios exactly one of these covariates induces DIF in one item. Again, all thresholds of one item are shifted in the same direction. There is one split-point $c_{vj}$ per item at $c_{vj}=x_{v_{med}}$. The threshold parameters of the two sub-groups are given by
       \[\gamma_{ir(2)}=\gamma_{ir(1)}+\lambda \cdot I(x_{pv}>x_{v_{med}}), \quad r=1,2.\]
			
To obtain weak, medium and strong DIF, parameters $\lambda$ are chosen in the same way as in the previous simulation.
	
\begin{figure}[!t]
\begin{center}
\includegraphics[width=0.48\textwidth]{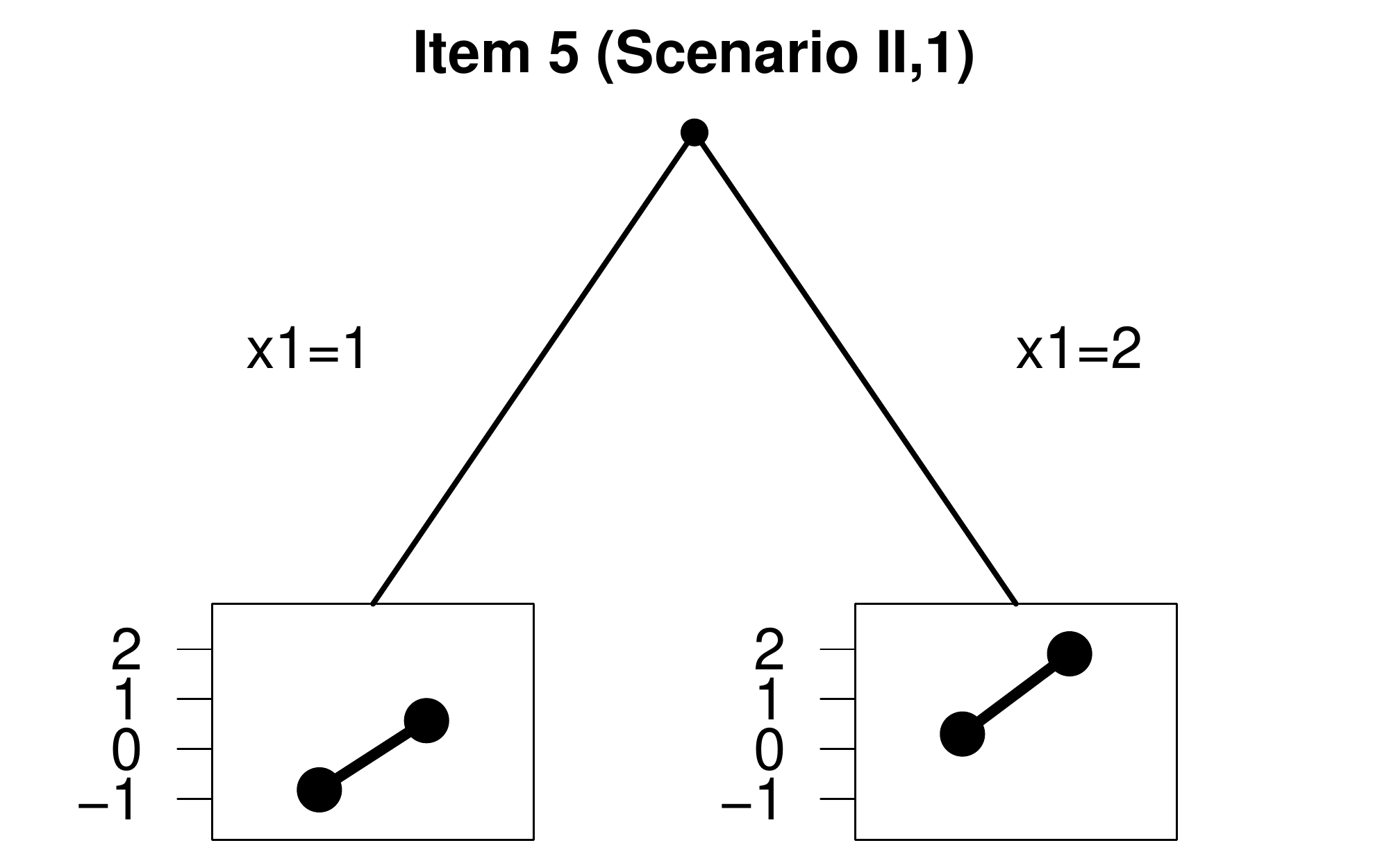}
\includegraphics[width=0.48\textwidth]{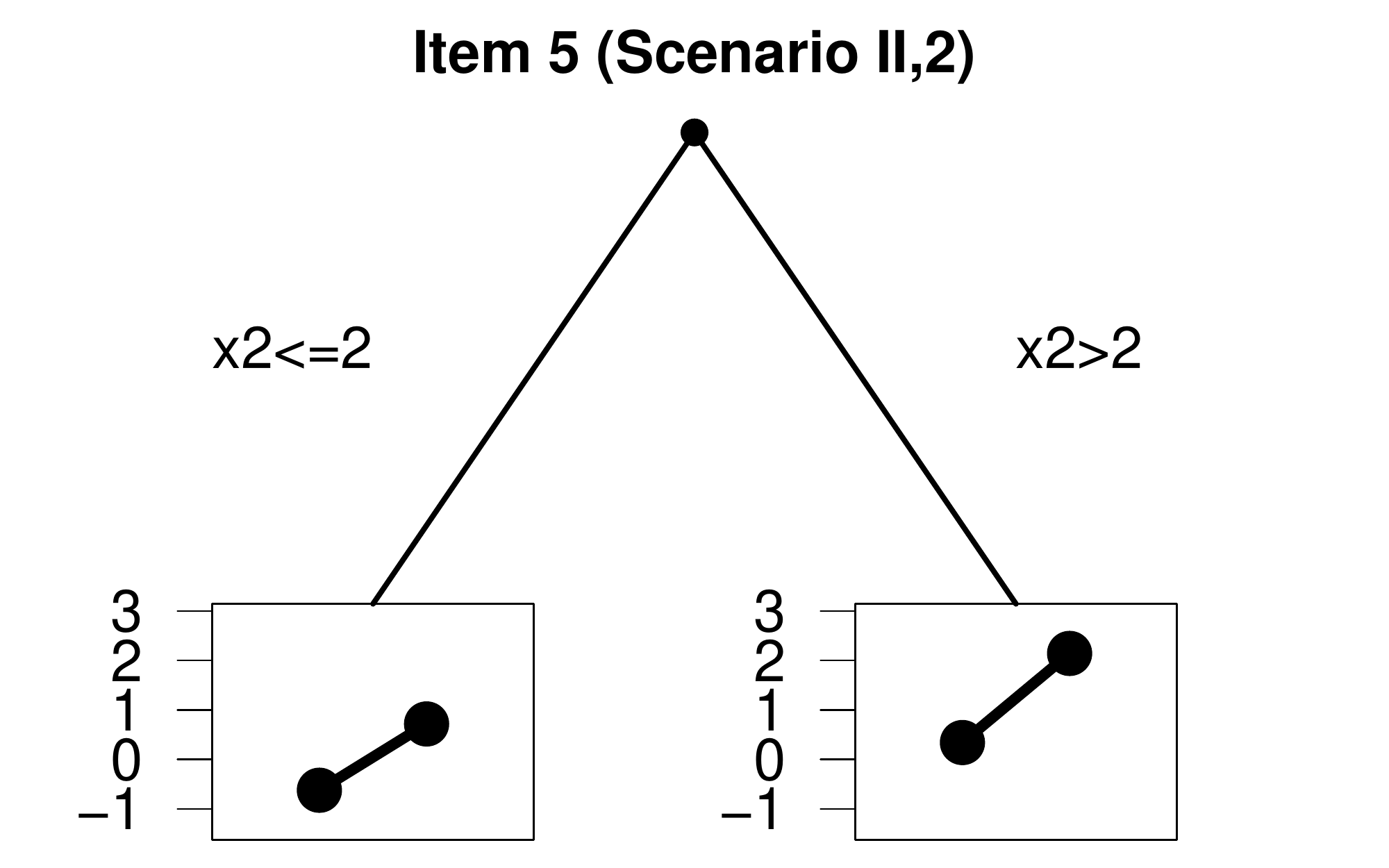}
\includegraphics[width=0.48\textwidth]{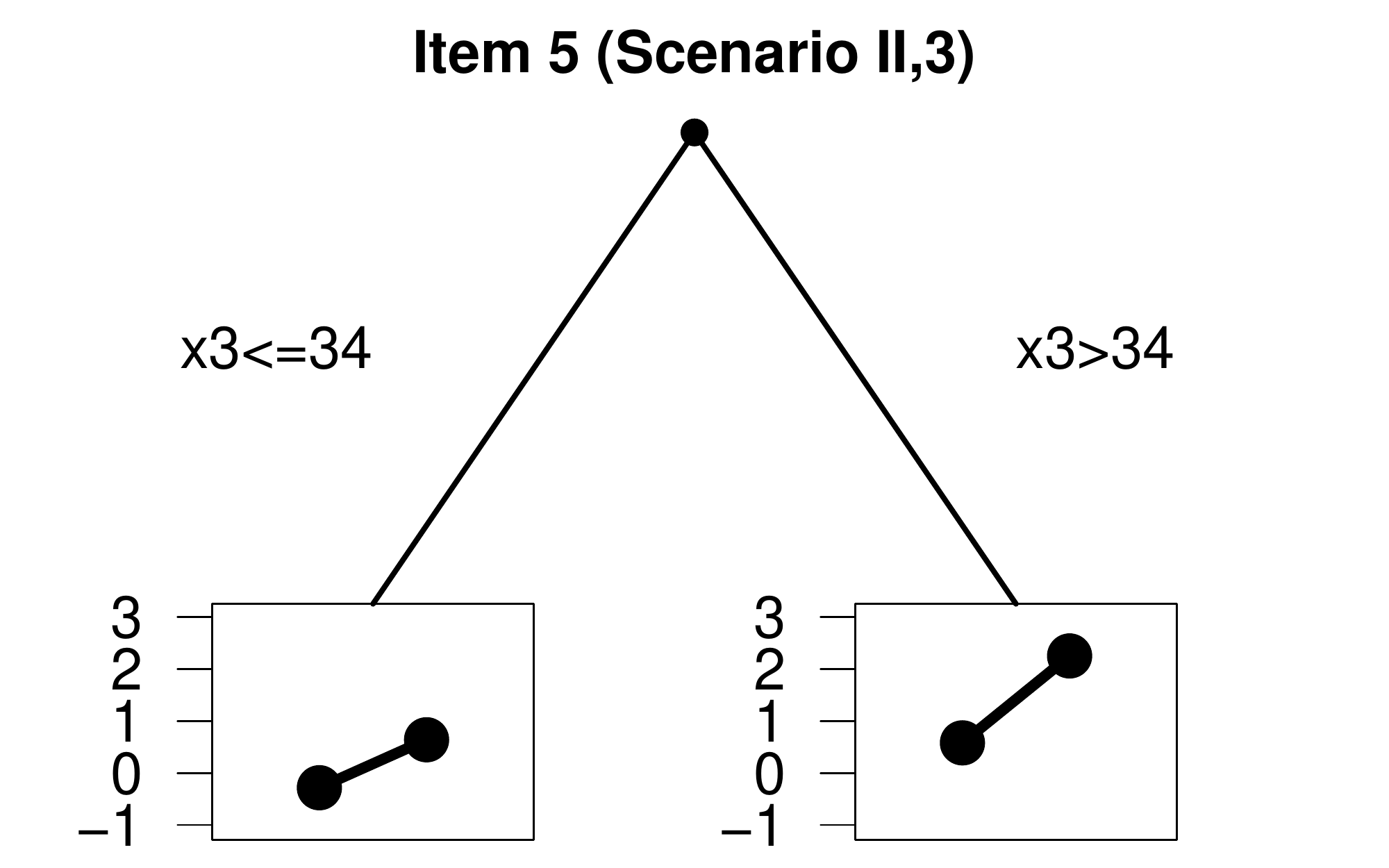}
\end{center}
\caption{Estimation results for one example of the three scenarios of Simulation II with three covariates and strong DIF. The estimated item parameters $\gamma_{5r(1)}$ and $\gamma_{5r(2)}$ are visualized in each leaf of the trees.}
\label{examplesII}
\end{figure}

\subsection*{Results }
		
Figure \ref{examplesII} shows one estimated tree for item 5 (the item with DIF) for the three different scenarios of Simulation II with strong DIF, respectively. In the chosen examples the true underlying DIF structure was detected. In scenario 1 DIF is induced by $x1$, in scenario 2 by $x2$ and in scenario 3 by $x3$. In these examples also the true simulated split-points ($2$, $2$ and $34$) are correctly identified. In each scenario, the true item parameters are $\gammab_{5(1)}=(-0.5,0.5)^\top$ in the left node and $\gammab_{5(2)}=(0.5,1.5)^\top$ in the right node. It can be seen from the graphical representations of the parameters in the leafs of the trees  that the estimated parameters are quite close to the true ones.

To account for the multiple covariates in the model the significance level at each node is divided by the number of covariates available at this node: $\alpha=0.05/V$. Table \ref{tab:2.simulation.ift} and Table \ref{tab:2.simulation.ift.alt} give an overview of the true and false positive rates based on 50 replications for Simulation II.

  \begin{table}[!t]
		\caption{ True positive and false positive rates for PCM-IFT (Simulation II)}
		\begin{center}
			\begin{tabularsmall}{llcccc}
				
				\toprule
       & \makebox{\textbf{DIF strength}}      & $\boldsymbol{TPR_I}$ & $\boldsymbol{FPR_I}$ & $\boldsymbol{TPR_{IV}}$ & $\boldsymbol{FPR_{IV}}$ \\
                \cline{2-6}
       Scenario 1 &     no DIF  & - & 0.048 & - & 0.028 \\
			&	weak  & 0.160 & 0.043 & 0.120 & 0.026 \\
			&	medium  & 0.556 & 0.044 & 0.533  & 0.036 \\
			&	strong  & 0.898 & 0.035 & 0.898 & 0.039 \\
            \cline{2-6}
	Scenario 2 &	no DIF  & -     & 0.045 & -	 & 0.027 \\
			&		weak	& 0.120 & 0.046 & 0.060 & 0.034 \\
			&		medium  & 0.306 & 0.047 & 0.306 & 0.028 \\
			&		strong  & 0.977 & 0.040 & 0.977 & 0.042 \\
          \cline{2-6}
	Scenario 3 & no DIF  & -	 & 0.048&  -	& 0.029\\	
            &    weak 	 & 0.102 & 0.055& 0.061 & 0.033 \\
			&	medium 	 & 0.630 & 0.040& 0.609 & 0.036 \\
			&	strong 	 & 1.000 & 0.045& 1.000 & 0.043 \\

				\bottomrule
			\end{tabularsmall}
		\end{center}
		
		\label{tab:2.simulation.ift}
	\end{table}

It can be seen in Table \ref{tab:2.simulation.ift} that false positive rates are always close to the given significance level demonstrating that the alpha level correction works quite well. For the combination of items and variables they are necessarily smaller.  From the first and the third column in Table \ref{tab:2.simulation.ift}, it can be seen that almost in all cases where a split was performed, also the right variable was selected. On variable level (Table \ref{tab:2.simulation.ift.alt}) false positive rates again are higher than 0.05 but not as high as in simulation I. In simulation II the split has to be made for the right variable and therefore all rates are divided by the number of covariates in the end. It's noteworthy that TREE-PCM is very conservative in this simulation which results in very small true and false positive rates. Similar to simulation I, true positive rates of PCM-IFT are much higher than those of TREE-PCM.

\begin{table}[!t]
		\caption{ $TPR_V$ and $FPR_V$ for TREE-PCM and PCM-IFT (Simulation II)}
		\begin{center}
			\begin{tabularsmall}{llcccc}
				
				\toprule
				& \makebox{\textbf{DIF strength}} & \multicolumn{2}{c}{\textbf{TREE-PCM}} & \multicolumn{2}{c}{\textbf{PCM-IFT}} \\
				&&$TPR_V$&$FPR_V$&$TPR_V$&$FPR_V$\\
                \cline{2-6}
     Scenario 1 & no DIF &---    & 0.007 &---& 0.127  \\
			&	weak 	 & 0.040 & 0.010 & 0.240 & 0.100   \\
			&	medium	 & 0.180 & 0.030 & 0.600 & 0.100   \\
			&	strong	 & 0.860 & 0.020 & 0.898 & 0.061   \\
            \cline{2-6}
	Scenario 2 &  no DIF &---& 0.007 &---&  0.120 \\	
            &    weak    & 0.040 & 0.010 & 0.120 & 0.160 \\
			&	medium	 & 0.100 & 0.010 & 0.340 & 0.110 \\
			&	strong	 & 0.700 & 0.020 & 0.977 & 0.093 \\
            \cline{2-6}
	Scenario 3 &	no DIF  &---	& 0.007 &---& 0.129 \\
			&	weak 		& 0.040 & 0.000 & 0.204 & 0.143 \\
			&	medium		& 0.100 & 0.020 & 0.652 & 0.098 \\
			&	strong		& 0.960 & 0.030 & 1.000 & 0.104 \\
				\bottomrule

			\end{tabularsmall}
		\end{center}
		
		\label{tab:2.simulation.ift.alt}
	\end{table}


		

 \subsection{Simulation III: Non-Homogenous DIF} \label{SimIII}

In the third simulation non-homogenous DIF is simulated in the settings with $I=8$, $k=3$ and $I_{DIF}=1$ with regard to one binary DIF inducing covariate.
Unlike in the previous simulations, threshold parameters now are not all shifted by an equal amount from the reference to the focal group, but half of the parameters is shifted to the left and the other half to the right. More precisely, since we only consider the case of two threshold parameters per item the first threshold parameter is shifted to the left and the second to the right. As a result, the difference between threshold parameters, i.e. the category width changes from the reference to the focal group. The two threshold parameters are then given through:
 \[\gamma_{i1(2)}=\gamma_{i1(1)}-\lambda \cdot I(x_p=1) \]
 \[ \gamma_{i2(2)}=\gamma_{i2(1)}+\lambda \cdot I(x_p=1). \]

 \begin{table}[!t]
		\caption{ True positive and false positive rates for PCM-IFT (Simulation III)}
		\begin{center}
			\begin{tabularsmall}{lcc}
				
				\toprule
        \makebox{\textbf{DIF strength}}      & $\boldsymbol{TPR_I}$ & $\boldsymbol{FPR_I}$ \\
                \cline{1-3}
           no DIF  & - & 0.065  \\
			weak   & 0.220 & 0.051 \\
			medium  & 0.560 & 0.057  \\
			strong  & 0.980 & 0.054  \\
				\bottomrule
			\end{tabularsmall}
		\end{center}
		
		\label{tab:3.simulation.ift}
	\end{table}

\begin{figure}[!t]
\begin{center}
\includegraphics[width=0.48\textwidth]{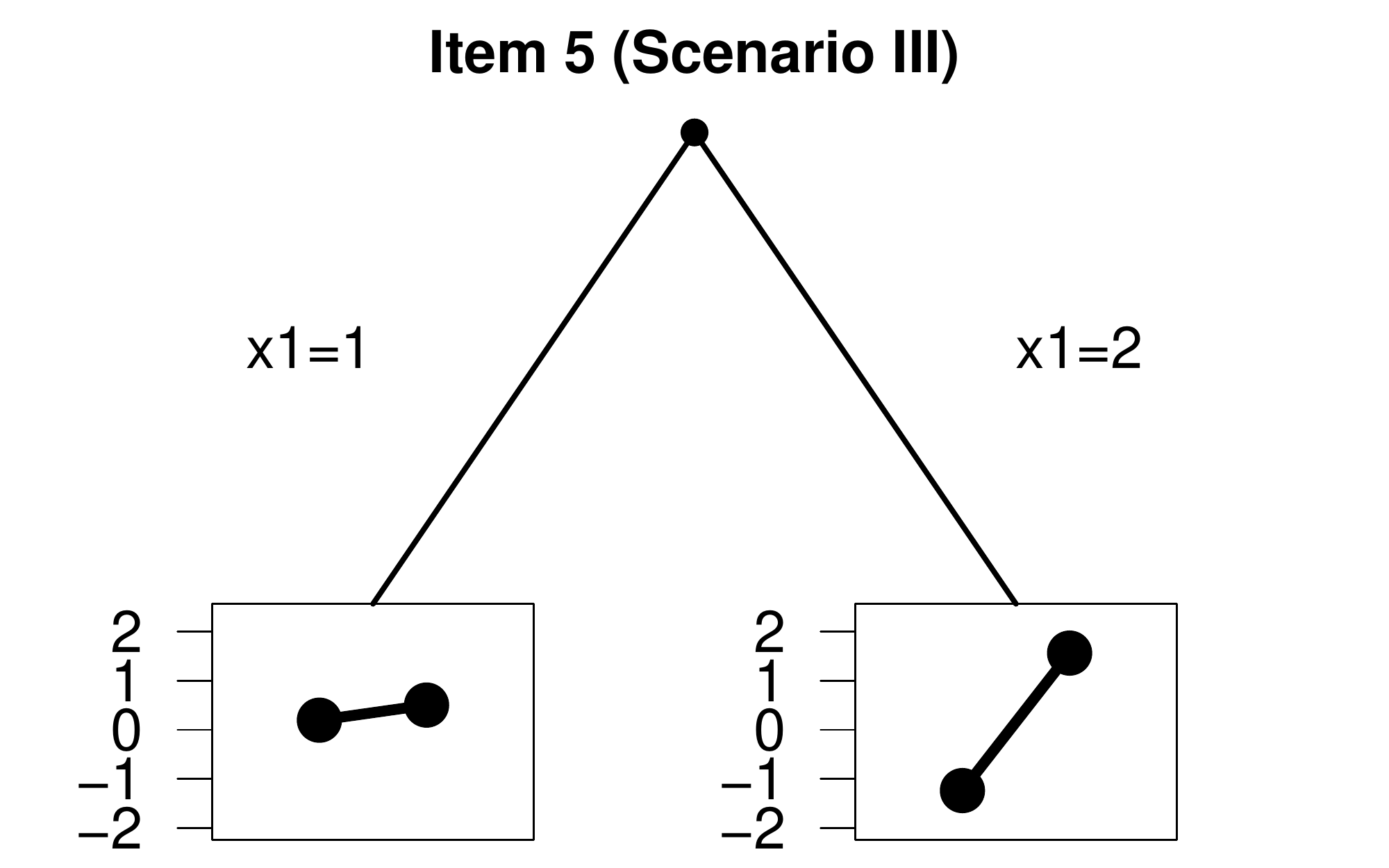}
\end{center}
\caption{Estimation result for one example of Simulation III with one covariate and non-homogenous DIF (strong setting). The estimated item parameters $\gamma_{5r(1)}$ and $\gamma_{5r(2)}$ are visualized in each leaf of the tree.}
\label{exampleIII}
\end{figure}

\subsection*{Results }

Table \ref{tab:3.simulation.ift} displays true and false positive rates on the item level for PCM-IFT. Both, false positive and true positive rates are satisfactory and very similar to those in simulation I, scenario 1 where the same number of items and categories were used.
Figure \ref{exampleIII} shows one estimated tree for item 5 (the item with DIF) for the setting with strong DIF of Simulation III, where the true underlying DIF structure was detected. In this scenario with non-homogenous DIF the true item parameters are $\gammab_{5(1)}=(-0.5,0.5)^\top$ in the left node and $\gammab_{5(2)}=(-1.5,1.5)^\top$ in the right node. It can be seen from the graphical representations of the parameters in the leafs of the trees, that the underlying non-homogeneous DIF structure is detected by the algorithm.

\section{Application}\label{sec:App}

	In this section, the new PCM-IFT approach is applied to real data. This allows us to draw conclusions about its functioning in real circumstances.
		We examine two facets from the major domain \emph{Openness to Experience} of the same data set that was used in the illustrative example in Section \ref{sec:ill.ex}. The whole test comprises 240 items that are answered on a Likert type scale from 0 \emph{strongly disagree} to 4 \emph{strongly agree}. One sub-facet comprises 8 items and 6 sub-facets in turn build one of the 5 major domains Neuroticism, Extraversion, Openness to Experience, Agreeableness and Conscientiousness.
	
\begin{figure}[!t]
			\centering
			\includegraphics[width=\textwidth]{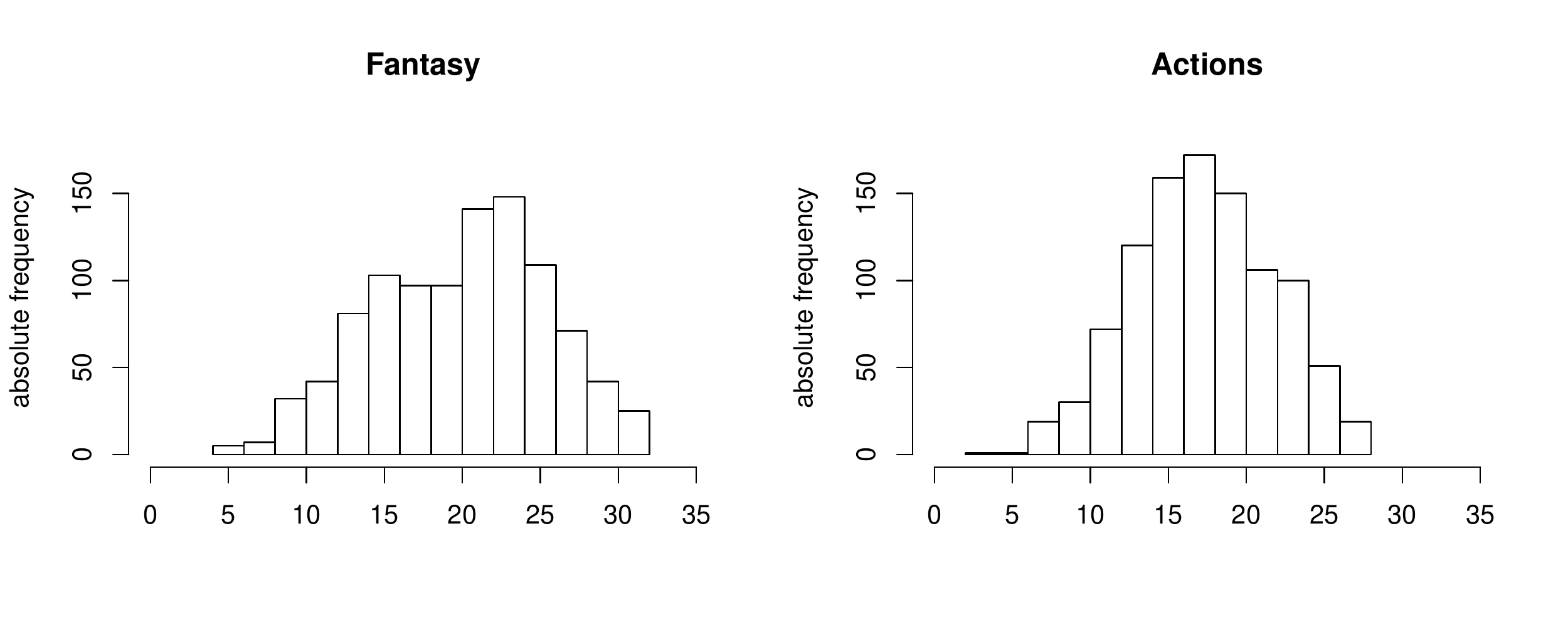}
			\caption{Graphical representation of the distribution of the sum scores of the facets \emph{Fantasy} and \emph{Actions} (NEO-PI-R).}
			\label{descriptives_facets}
		\end{figure}

	The major domain \emph{Openness to Experience} is described in the manual as \emph{the active seeking and appreciation of experiences for their own sake}. Here we analyse the two sub-facets \emph{Fantasy} (\emph{receptivity to the inner world of imagination}) and \emph{Actions} (\emph{openness to new experiences on a practical level}) \citep{ostendorf_neo-personlichkeitsinventar_2004}.  Each of the items again has five categories. Distributions of the sum scores of the two sub-facets are displayed in Figure \ref{descriptives_facets}.

	    	To test for DIF in the two facets we incorporate the covariates gender (male: 1, female: 2) and age.
			 The distributions of the covariates can be obtained from Figure \ref{descriptives_C4}.  The 1000 subjects were randomly drawn out of the 11,724 cases of the norm data set. The sample comprises 4216 males and 7498 females with age ranging from 16 to 91.
	

	    \subsection{PCM Item-Focussed Trees}

	    For the sub-facet \emph{Fantasy}, two of the eight items were diagnosed as  DIF items.  The two items with DIF were the following:
	
	    	\begin{itemize}
			\item[] Item 3: I have an active and lively fantasy life.
			\item[] Item 6: When I feel that my thoughts are drifting off into daydreams I usually become busy and start to focus on a task or an activity. (R)
	    	\end{itemize}
		The (R) behind item 6 indicates that this item was reverse coded. This means that strong agreement to this question indicates a low level of fantasy. For simplicity, all items that are reverse coded have been recoded before the analysis. Therefore, for all analyses a high value on this item means the person disagreed to the question.

\begin{figure}[!t]
			\centering
			{\includegraphics[width=\textwidth]{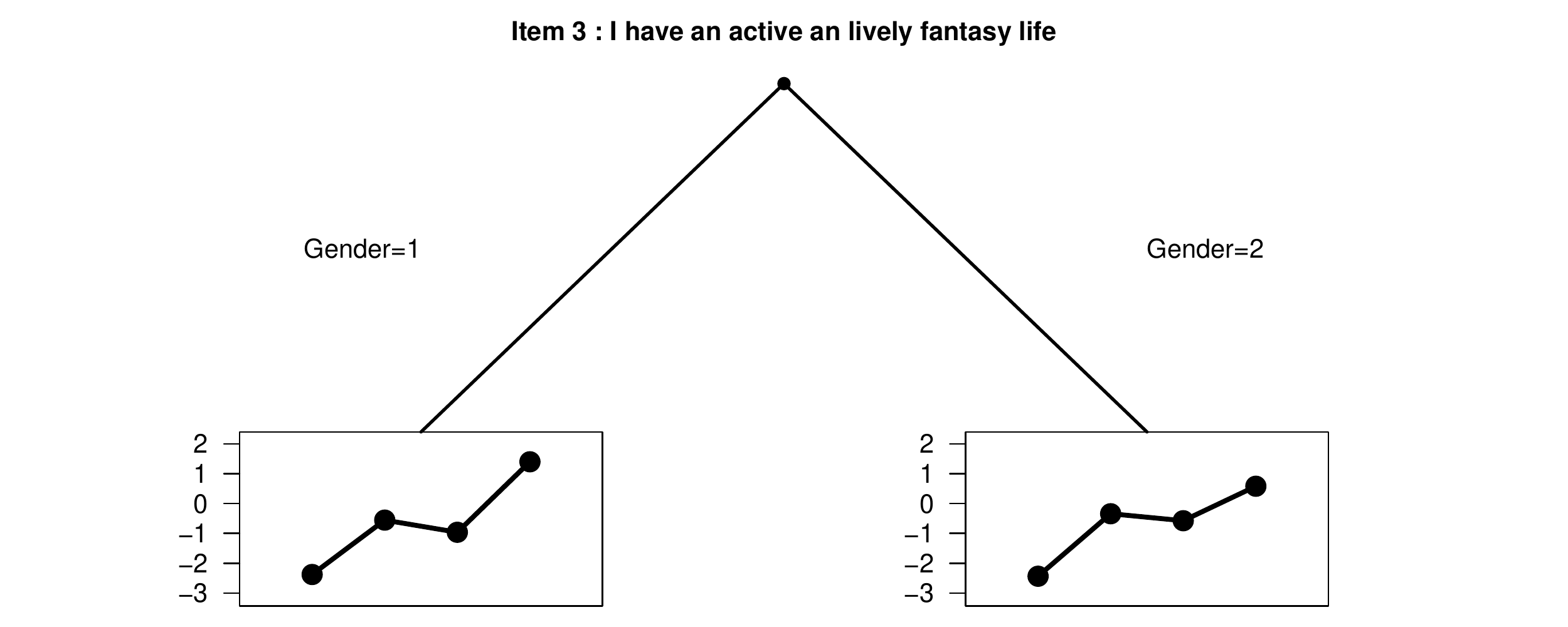}}
	
	        \vspace{1cm}
	
			{\includegraphics[width=\textwidth]{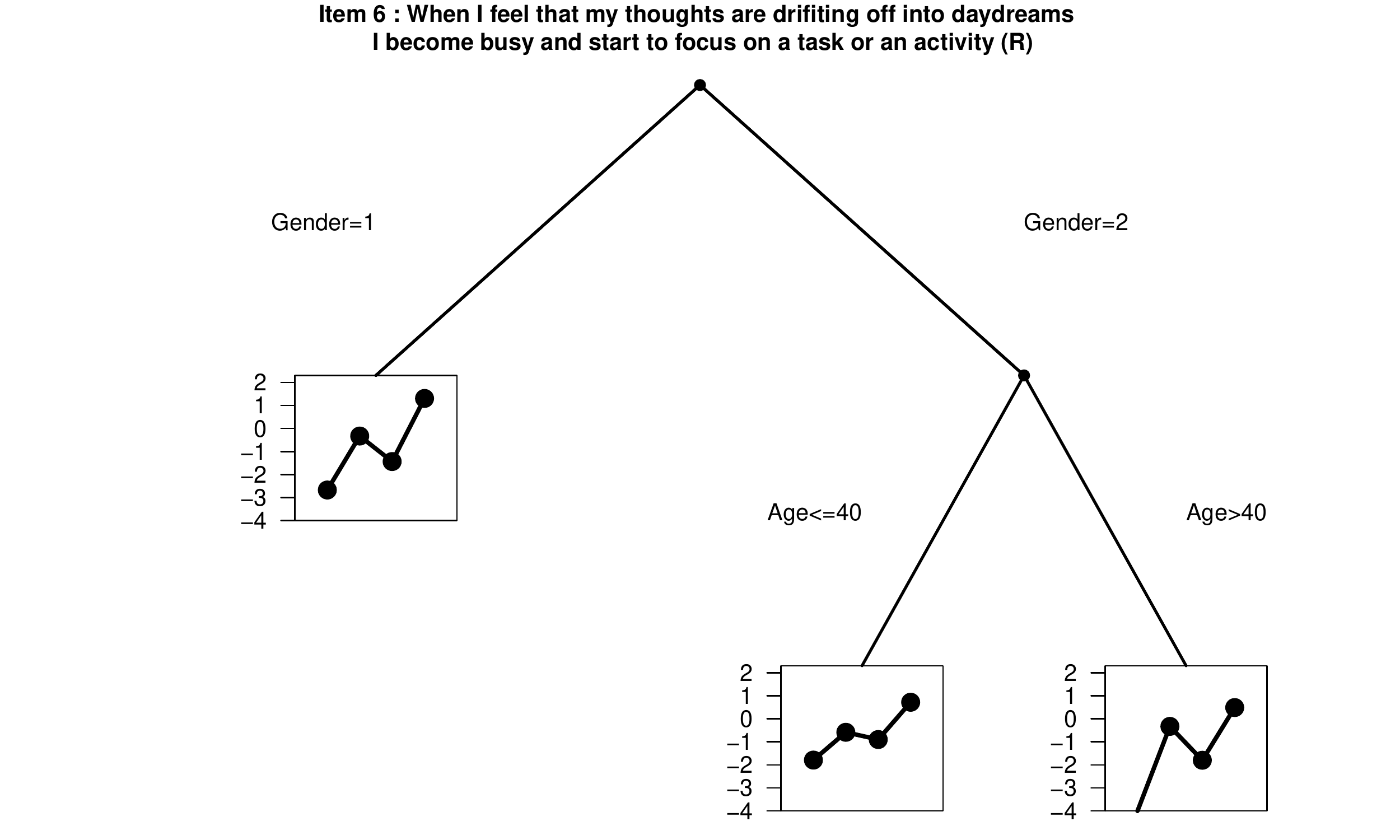}}
	\caption{Trees for item 3 and item 6 of the sub-facet \emph{Fantasy} (NEO-PI-R).}
			\label{tree_O1}
		\end{figure}

Item 3 was only split once in covariate gender. The resulting tree is shown in Figure \ref{tree_O1} (upper panel). At the terminal nodes the four threshold parameters for the respective partition are given.  It is seen that for both groups thresholds are not ordered indicating that a higher latent trait is required for passing the second threshold than for passing the third threshold. This effect is slightly more extreme for males than for females. Also, for males an even higher latent trait is required to pass the fourth threshold.
	
	    Item 6 was split twice with regard to gender and age. The first split was found for variable gender and within the the sub-group of females it is distinguished between younger women ($Age \leq 40$) and older women ($Age>40$).
		The resulting tree is shown in the lower panel of Figure \ref{tree_O1}.
	    It is seen that in a similar way as for item 3, in none of the terminal nodes the thresholds are ordered. The main difference between the three groups is the variation of the threshold parameter $\delta_{61}$, which is highest for females with age $\leq 40$ and lowest for females with age $>40$. For the latter this threshold parameter was even below $-4$ and is therefore not visible in the Figure anymore because the plot is truncated at $-4$. Since, the item is reverse coded, this means that for older females the probability was particularly low to pass the last threshold from \emph{agree} to \emph{strongly agree} for this question. A look at the answers shows that in this group (terminal node 3) only 2 persons (out of 133) had chosen the last category.
			
						    \begin{figure}[p]
	    		\centering
	
	           {\includegraphics[width=\textwidth]{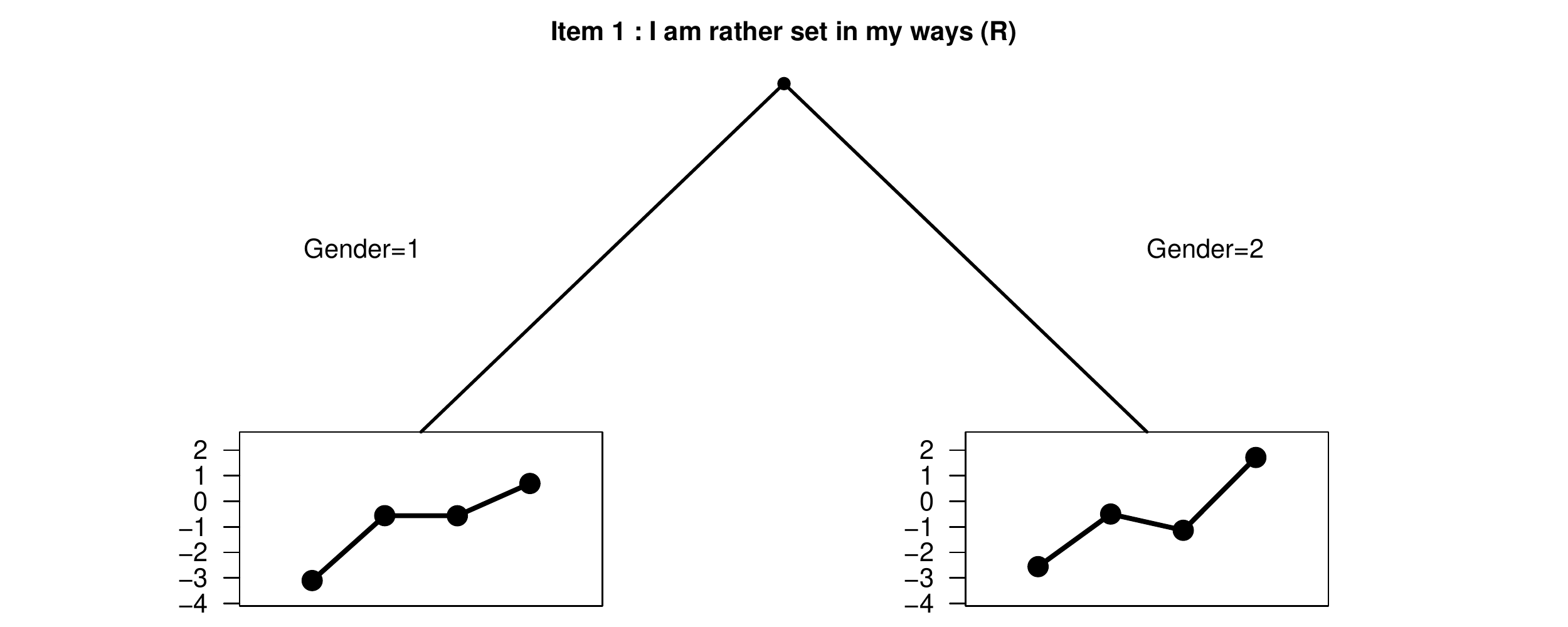}}
	

	           {\includegraphics[width=\textwidth]{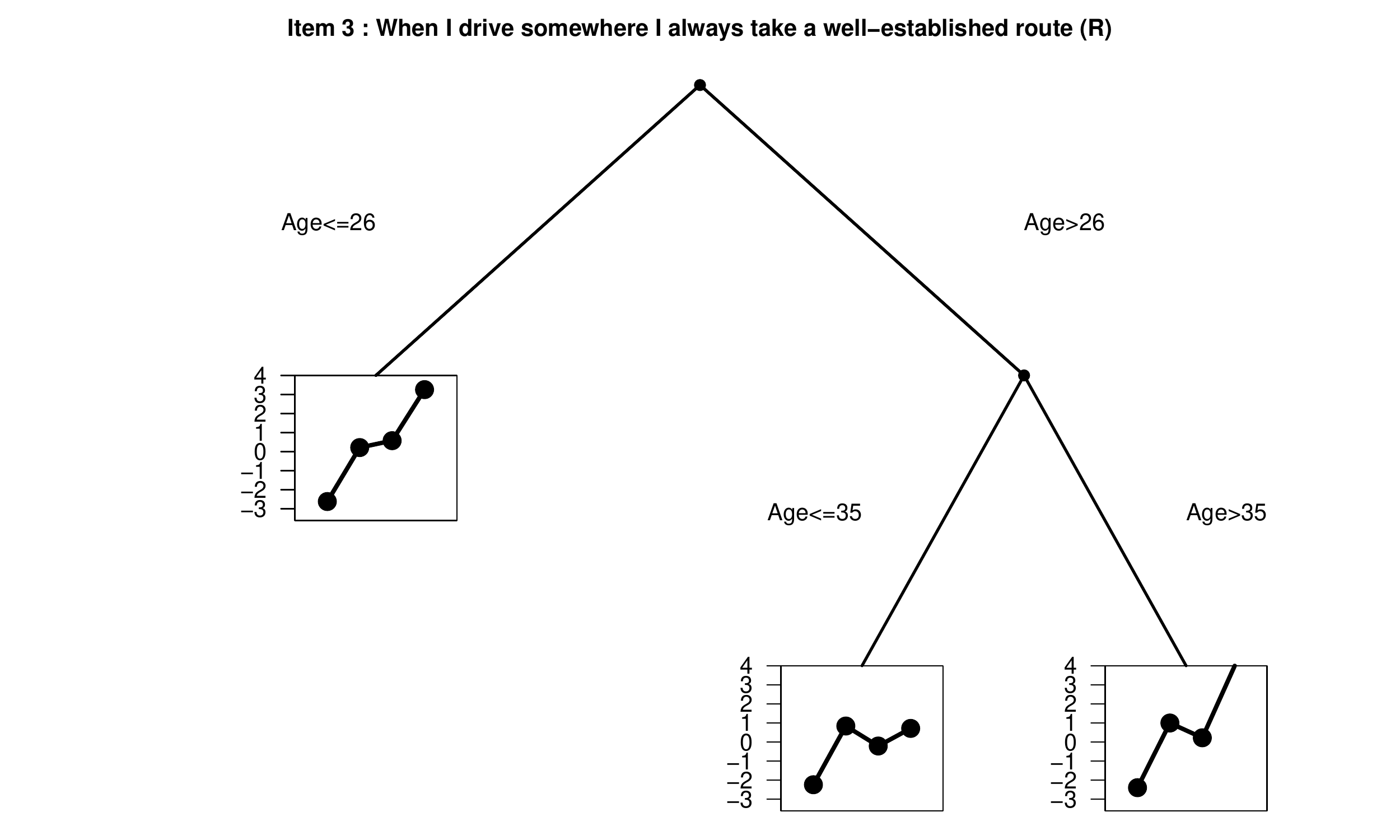}}
	
	
	    	        {\includegraphics[width=\textwidth]{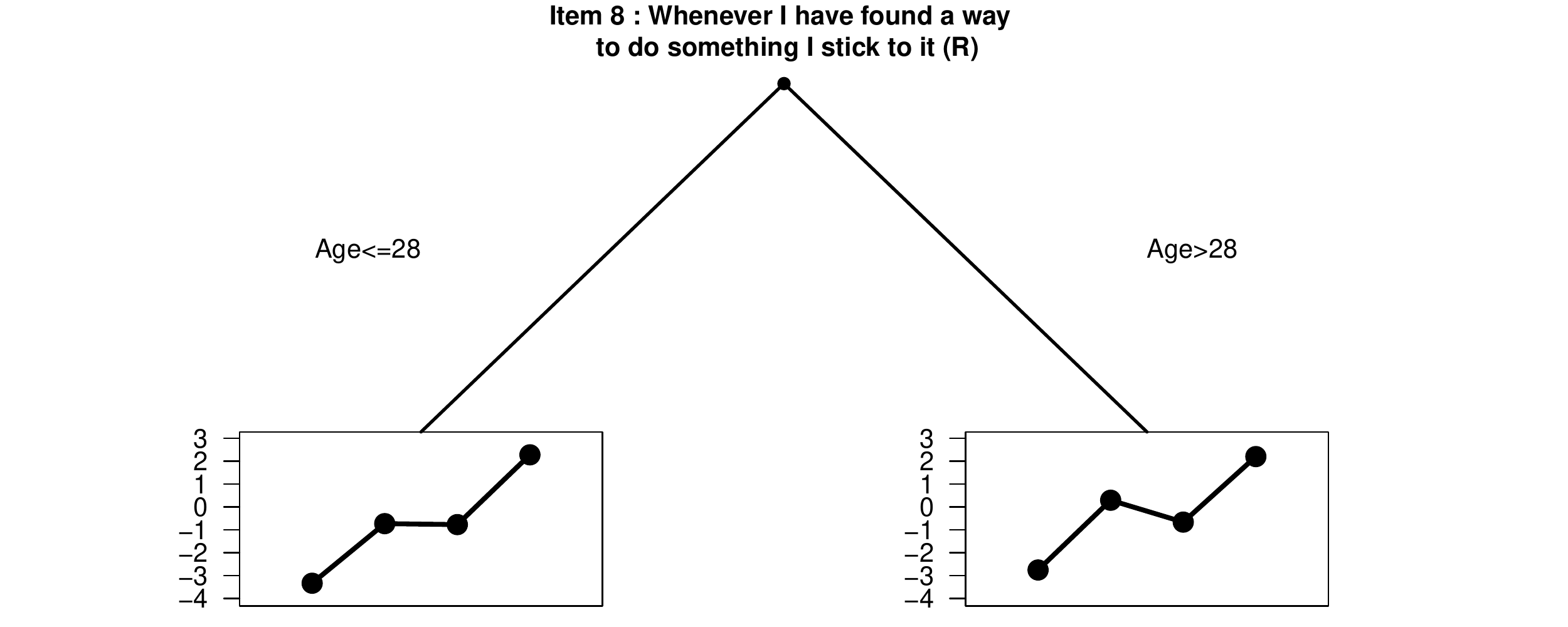}}
	    	
	    		\caption{Trees for the Items 1, 3 and 8 of the facet \emph{Actions} (NEO-PI-R).}
	    		\label{tree_O4_I3}
	    	\end{figure}
	
	     For the sub-facet \emph{Actions}, three items were detected to have DIF. Altogether the algorithm performs four splits until further splits are not significant anymore. The three items that have DIF are the following:
	
	    	\begin{itemize}
	    		\item[] Item 1: I am rather set in my ways. (R)
	    		\item[] Item 3: Whenever I have found a way to do something I stick to it. (R)
	            \item[] Item 8: When I drive somewhere, I always take a well-established route. (R)
	
	    	\end{itemize}
	    Item 1 was split for variable gender. Item 3 was split twice for the covariate age and item 8 was split once for the covariate age. 	
	    The trees for these items can be obtained from Figure \ref{tree_O4_I3}. Because of their reverse coding (R), these three items are all recoded for the analyzes. This means, that higher categories stand for lower agreement to the question.

	    The tree for item 1 shows that thresholds $\delta_{11}$ and $\delta_{14}$ are a little higher for females than for males. That means that in order to jump from category 0 to 1 as well as from 4 to 5, females need higher person parameters.  For item 3, the threshold estimates for $\delta_{34}$ differ most between the three groups. For persons over 35 years of age this parameter in fact yields a value of 18.729 and is therefore not visible because the plot is truncated at $4$. Thus, a particularly high latent expression is required to pass the last threshold which means to go from \emph{disagree} to \emph{completely disagree}. A similar pattern can be found in the group $Age \leq 26$. In both groups parameters of thresholds 1 and 4 are very far apart. This means, people rather have a tendency to the middle categories than to extreme categories in these two groups. The tree for item 8 shows that threshold parameters are not ordered for persons over 28 years of age.
	
		    \begin{figure}[!t]
	    \centering
	    \includegraphics[width=0.9\textwidth]{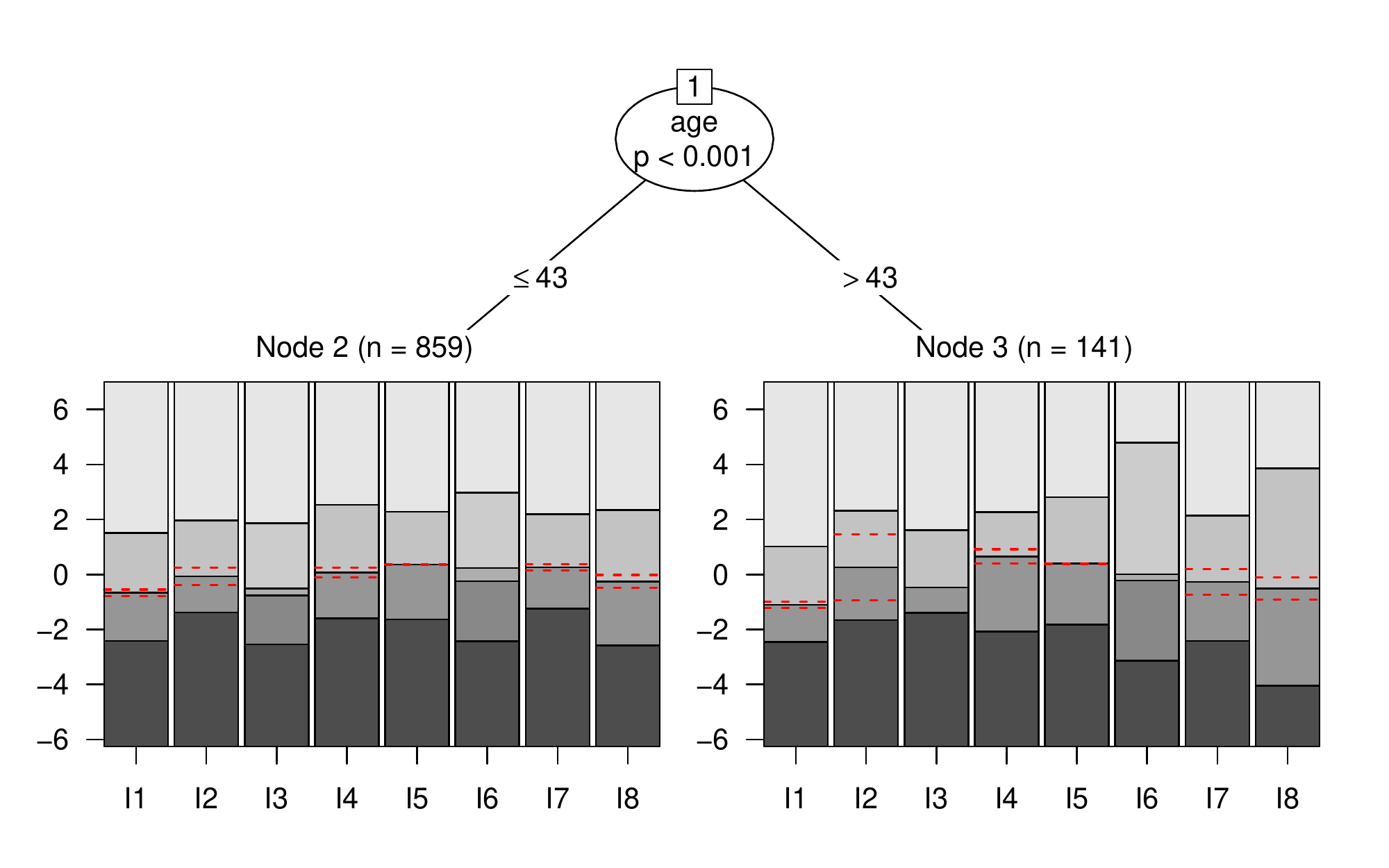}
	    \caption{Estimated tree for TREE-PCM of the sub-facet \emph{Fantasy} (NEO-PI-R).}
	    \label{pcmtree_plot_O1}
	    \end{figure}
			
				 \begin{figure}[!t]
		    \centering
		    \includegraphics[width=0.9\textwidth]{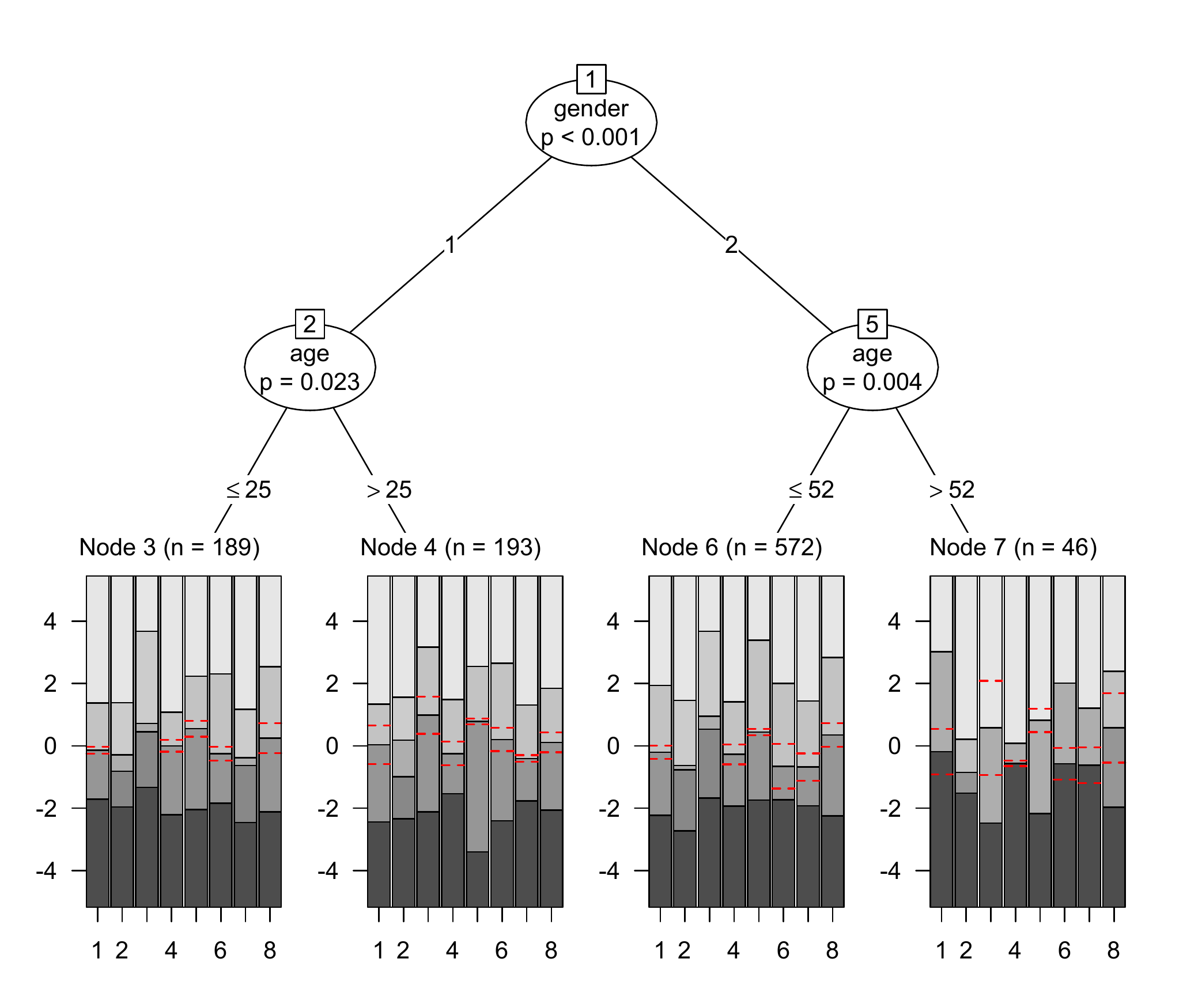}
		    \caption{Estimated tree for TREE-PCM of the sub-facet \emph{Actions} (NEO-PI-R).}
		    \label{pcmtree_plot_O4}
		    \end{figure}

		\subsection{Partial Credit Tree Approach}
		
	    \label{sec:PCtree}
	
	To illustrate the difference between the PCM-IFT approach and the partial credit tree (TREE-PCM) proposed by \cite{el2014detecting} we  analyze the same data sets also by using TREE-PCM.
		We use the same significance level as for PCM-IFT, namely $\alpha=0.05$.
		The resulting models for the sub-facet \emph{Fantasy} and the sub-facet \emph{Actions} are presented separately in Figures \ref{pcmtree_plot_O1} and \ref{pcmtree_plot_O4}.
		
		For the sub-facet \emph{Fantasy}, it shows only one split for the variable age at 43 years of age. At each terminal node an effect plot is shown for each item. The effect plot displays regions of most probable category responses over the range of the latent trait i.e. the regions between two adjacent thresholds. If two thresholds are reversed, the intermediate region is not shown but is indicated by horizontal dashed lines.
	According to PCM-IFT, the null hypothesis of one joint PCM has to be rejected for the facet \emph{Fantasy} since there is more than one terminal node. In contrast to the results of PCM-IFT, the partial credit yields ordered thresholds for items 3 and 6. Nevertheless, these two items do reveal strong differences in the effect plots between the two groups. However, from this plot it is not easy to identify the items that are responsible for DIF in this sub-facet because almost all items show light to strong differences in the plots between the two groups.
	
The two methods agree on age being a DIF inducing covariate for this facet. However, only PCM-IFT also identifies gender as DIF inducing variable.
It is not surprising that the results show differences. TREE-PCM uses a global strategy, after a split into age groups the overall differences of further splits are not strong enough to warrant further splits. In contrast, PCM-IFT uses an item-focussed strategy. It performs splits if differences between groups are large for specific items. For the two items the differences were strong enough in gender groups although they were not so strong on the global level to yield a split when using TREE-PCM. For TREE-PCM the dominating split was found in age. By construction PCM-IFT method is  more sensitive to DIF in only a few items while TREE-PCM is more sensitive to DIF in multiple items.


	For the sub-facet \emph{Actions} three splits were executed resulting in 4 terminal nodes. First, it was split for gender and then both nodes were split again for covariate age. In the sub-group of males it is split at 25 years of age, for females it is split at 52 years of age. It is not directly visible from the four parameter plots, which items are responsible for the DIF in the four groups. Moreover, a direct comparison of the results of this approach to the new PCM-IFT approach is not straightforward since only one tree was built for all items together while for PCM-IFT we obtained three different trees with different split-points. Nevertheless, there is some accordance since both techniques detect both covariates as DIF inducing covariates.

\section{Concluding Remarks}

We propose an approach to detect DIF in ordinal item response based on the partial credit model. By item-focussed recursive partitioning the proposed method allows for simultaneous detection of items and variables that are responsible for DIF. The results are small trees for each item that is not compatible with the PCM. Graphical representations of the threshold parameters in each terminal node enable  an easy interpretation of the estimated effects and the differences between the detected groups. The simulations demonstrate that the procedure works well, in particular in settings where only few DIF items with small DIF effects are present (which is usually the case in applications).

The proposed model explicitly tests DIF on the \emph{item level}. That means in each step the whole parameter vector ($H_0:\gammab_{i(1)}-\gammab_{i(2)}=\0$) is tested and if a split is performed all the threshold parameters are estimated in both nodes without any restrictions. An alternative strategy would be to test for DIF in single thresholds. Then for fixed item and variable in each step one tests the hypotheses $H_0:\gamma_{ir(1)}-\gamma_{ir(2)}=0,\; r=1,\hdots,k,$ and selects the threshold that has the best fit. Accordingly, in each step only one threshold $\delta_{ir}$ changes for one group. In future research one might also consider a \emph{homogeneous modelling approach}, in which again all thresholds are shifted but now all in the same direction by an item-specific constant $\gamma_i$. Then, for example, after the first split the item parameters in region $\{x_{pv}>c_v\}$ are defined by $\delta_{i1}+\gamma_i,\hdots,\delta_{ik}+\gamma_i$. Both strategies are certainly worth investigating but the adoption of the existing procedure needs further research.

We restricted  consideration to the widely used partial credit model. However, the basic concept can also be used to model DIF in alternative ordinal item response models, for example in the rating scale model (RSM; \citealp{Andrich:78}). In the RSM the predictor has the form $\theta_p-(\beta_i+\tau_r)$, with item location parameter $\beta_i$ and threshold parameter $\tau_r$. With item-focussed trees the location parameter $\beta_i$ can be replaced by $\gamma_{i(1)}I(x_{pv}\leq c_v)+\gamma_{i(2)}I(x_{pv}>c_v)$, the threshold parameter $\tau_r$ can be replaced by $\alpha_{r(1)}I(x_{pv}\leq c_v)+\alpha_{r(2)}I(x_{pv}>c_v)$ or both parameters can be modified simultaneously. Fitting of corresponding models requires the development of tailored testing strategies and appropriate estimation tools which is beyond the scope of this article.

All the results presented in this article were obtained by an R program (as described in Section \ref{computation}) which is available from the authors and will soon be available in an add-on package on CRAN.

\bibliography{literatur}

\end{document}